\documentclass{egpubl}
\usepackage{eurovis2025}

\SpecialIssuePaper         

\CGFccby

\usepackage[T1]{fontenc}
\usepackage{dfadobe}  

\usepackage{cite}  
\BibtexOrBiblatex
\electronicVersion
\PrintedOrElectronic
\ifpdf \usepackage[pdftex]{graphicx} \pdfcompresslevel=9
\else \usepackage[dvips]{graphicx} \fi

\usepackage{egweblnk}

\usepackage{amsmath}

\usepackage{hyperref}
\usepackage{multirow}
\usepackage{graphicx}
\usepackage{subcaption}
\usepackage{tikz}
\usepackage{xcolor}
\usepackage{float}
\usepackage{amsmath,amssymb}

\definecolor{bluefill}{HTML}{DAE8FC}
\definecolor{blueborder}{HTML}{6C8EBF}
\definecolor{redfill}{HTML}{F8CECC}
\definecolor{redborder}{HTML}{B85450}

\newcommand{\blueletter}[1]{%
  \tikz[baseline=(char.base)]{
    \node[shape=circle, draw=blueborder, fill=bluefill, text=black, line width=1pt, inner sep=1pt] (char) {\scalebox{0.9}{#1}};  }%
}
\newcommand{\redletter}[1]{%
  \protect\tikz[baseline=(char.base)]{
    \protect\node[shape=circle, draw=redborder, fill=redfill, text=black, line width=1pt, minimum size=1.1em, inner sep=0pt] (char) {\protect\scalebox{0.8}{#1}};  }%
}

\def\name{ViBEx}
\def\longname{Visual Bias Explorer}

\newcommand{\rev}[1]{{\color{black}{#1}}} 

\title[\longname]%
      {Interactive Discovery and Exploration of Visual Bias in\\ Generative Text-to-Image Models}

\author[J. Eschner et al.]
{\parbox{\textwidth}{\centering 
        Johannes Eschner$^1$\orcid{0009-0001-6784-8503},
        Roberto Labadie-Tamayo$^{2}$\orcid{0000-0003-4928-8706}, 
        Matthias Zeppelzauer$^{2}$\orcid{0000-0003-0413-4746}
        and Manuela Waldner$^1$\orcid{0000-0003-1387-5132}  
        }
        \\
{\parbox{\textwidth}{\centering 
        $^1$TU Wien, Austria\\
        $^2$St. Pölten University of Applied Sciences, Austria
       }
}
}

\begin{document}

\teaser{
 \includegraphics[width=0.86\linewidth]{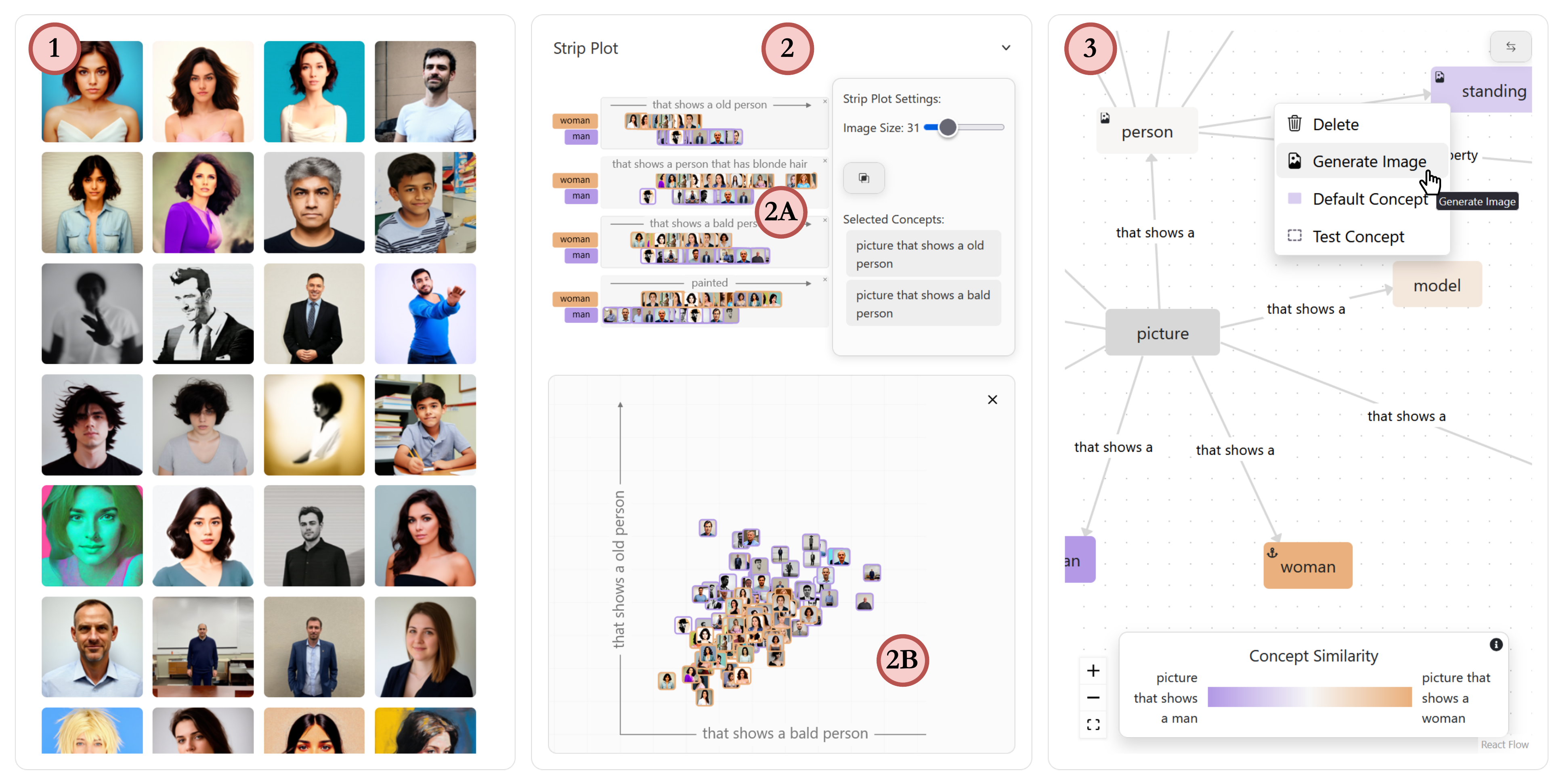}
 \centering
    \caption{\name~interface: \textbf{data view} \redletter{1} with a collection of generated images corresponding to concepts in \redletter{3}. \textbf{Bias visualizations}:  \redletter{2} \textbf{strip plots} \redletter{2A} showing image-to-text similarities for selected test concepts and \textbf{intersectional bias scatterplot} \redletter{2B} showing the bivariate distribution of two intersected test concepts. \textbf{Prompting tree} \redletter{3} for externalization of approximate visual bias and as a prompting interface.}
    \label{fig:interface}
}

\maketitle
\begin{abstract}
   Bias in generative Text-to-Image (T2I) models is a known issue, yet systematically analyzing such models' outputs to uncover it remains challenging. We introduce the \longname~(\name) to interactively explore the output space of T2I models to support the discovery of visual bias. \name~introduces a novel flexible prompting tree interface in combination with zero-shot bias probing using CLIP for quick and approximate bias exploration. It additionally supports in-depth confirmatory bias analysis through visual inspection of forward, intersectional, and inverse bias queries. \name~is model-agnostic and publicly available. In four case study interviews, experts in AI and ethics were able to discover visual biases that have so far not been described in literature. 


\end{abstract}  
\section{Introduction}
Generative Text-to-Image (T2I) models have been shown to amplify demographic stereotypes and perpetuate social biases in their outputs~\cite{naik_social_2023, luccioni_stable_2023, wang_t2iat_2023, vice_quantifying_2023, bianchi_easily_2023}. Prior work has, for instance, observed biases in the depiction of gender and ethnicity across various occupations~\cite{wan_survey_2024}. Traditional approaches to bias exploration of T2I systems mostly focus on statistically analyzing known or anticipated biases~\cite{wang_t2iat_2023, dinca_openbias_2024, boratto_measuring_2023}. Similarly, bias mitigation methods focus on reducing pre-identified biases in T2I outputs~\cite{nichol_dalle_nodate, dinca_openbias_2024}. Next to mitigating already discovered biases, the discovery and thorough documentation of previously unknown bias dimensions is essential for creating fairer and more balanced models. Automated methods may uncover new bias dimensions~\cite{dinca_openbias_2024, chinchure_tibet_2024} but may also reveal unrelated entanglements, miss certain biases, or identify biases differently than a human observer. Therefore, we argue that human oversight is essential to determine what constitutes a bias, especially when it comes to different socio-cultural contexts.

Visual Analytics (VA) systems for interactive bias exploration mainly exist in the context of word embeddings~\cite{ghai_wordbias_2021} and text representations~\cite{kabir_stile_2024, hoque2022dramatvis}. For T2I models, however, there is a gap when it comes to the interactive bias analysis and discovery of new bias, as previous work focuses either on fully automatic and non-interactive pipelines~\cite{dinca_openbias_2024, liu_organizing_2024} or on interactively showcasing known biases~\cite{luccioni_stablediffusionbiasexplorer_2023}. This gap may be rooted in the difficulty of providing an interactive system given the higher computational demands of image generation systems over text models.

In this work, we focus on how bias manifests visually in AI-generated images.
Our aim is to support both the \emph{exploration} of potentially biased concepts to discover unexpected bias and the more in-depth \emph{confirmation} of expected or observed visual bias. To this end, we present a workflow for interactive bias discovery together with a reference implementation proposing tailored visualization components to showcase bias as well as performance and (expert) user studies of our approach.
Based on this workflow, we propose \name, a novel visual bias exploration approach to support the discovery of bias through interactive exploration and confirmation. \name~enables efficient, flexible analysis through several unique interface components. 
We discuss and show how these components ensure that \name~meets the key requirements for exploratory visual analysis systems, such as flexibility and real-time interactivity. 
In summary, our main contributions are:

\begin{enumerate}
    \item The conceptualization of a user-driven visual bias exploration workflow (\name), incorporating a novel interaction and visualization design to facilitate flexible, trustworthy, and rapid exploration and confirmation of biases in T2I systems. 
    \item A first model-agnostic reference implementation of the \name~workflow as a publicly available online resource to inspect visual bias in a real-time web application: \url{https://vibex.jde.cg.tuwien.ac.at} 
    \item A set of previously unidentified visual biases in Stable Diffusion 3 (SD3)~\cite{esser_scaling_2024}. These were discovered in case studies by experts in the fields of AI, media ethics, and digital humanism using \name~and validated via established bias quantification.  
\end{enumerate}

\section{Background and Related Work}
\label{sec:related}

We follow the formal bias definition by D'Incà et al.~\cite{dinca_openbias_2024} who consider a model as \emph{unbiased} if, in a class-agnostic context $t$, the set of classes $C$ exhibit a uniform distribution. Conversely, a generator is \emph{biased} if it is more likely to produce content of one class $c_i \in C$ (e.g., \emph{``man''}) given a neutral prompt $t$ (e.g., \emph{``A picture of a doctor''}). In our work, we assume that users explore potential bias with respect to \emph{concepts}. We will refer to the classes $C$ to be inspected as \emph{anchor concepts} and to the context $t$ as \emph{test concept}. \textit{Intersectional} bias occurs when biases associated with multiple test concepts interact to create unique representational harms~\cite{cabrera_fairvis_2019, tan_assessing_2019}.  
Biases in T2I outputs are grounded in social harms (e.g. under- or misrepresenting groups)~\cite{wan_survey_2024}, while entanglements are \emph{incidental correlations} without inherent harm~\cite{chinchure_tibet_2024}. For example, the concept \emph{``man''} may be predominantly depicted with a beard. 
We argue that, ultimately, it should be the user who decides what constitutes a bias versus an expected entanglement.

\subsection{Bias Measures and Mitigation}
\label{sec:biasMeasuresMitigation}

Bias evaluation in T2I systems generally involves creating curated datasets and applying bias evaluation metrics. Curated datasets are designed to target specific biases by carefully selecting prompts or using datasets like FairFace~\cite{karkkainen_fairface_2021}, Flickr 30k~\cite{young2014image} or MS COCO~\cite{lin2014microsoft}, which provide image captions suited for demographic or contextual prompts. Naik and Nushi~\cite{naik_social_2023} highlight social biases using demographic prompts, while others compare output distributions to training datasets, revealing amplification or reversals of biases~\cite{friedrich_fair_2023}.
Typical bias measures include classification-based methods, such as analyzing feature frequencies using vision-language models like CLIP (Contrastive Language-Image Pretraining)\cite{radford_learning_2021} or Visual Question Answering (VQA)\cite{zhang_auditing_2023}, and embedding-based approaches that assess association scores in embedding space~\cite{boratto_measuring_2023, wang_t2iat_2023}.

Prior work on bias measures for T2I systems mainly focuses on \emph{gender}, \emph{race}, and \emph{skin tone}. Luccioni et al.~\cite{luccioni_stable_2023} find that popular T2I systems underrepresent marginalized identities. They highlight these biases by comparing profession-based image prompts to U.S. labor statistics. Similarly, Naik and Nushi~\cite{naik_social_2023} show that DALL·E 2~\cite{ramesh_zero-shot_2021} and Stable Diffusion v1~\cite{rombach_high-resolution_2022} amplify social biases in gender, ethnicity, age, and geography, often depicting underrepresented regions in adverse conditions. Recent work expands these known biases by incorporating appearance attributes (e.g., grooming, accessories)~\cite{liu_organizing_2024} and factors like activity, object size, and emotion~\cite{dinca_openbias_2024}.

To minimize the presence of known biases in T2I outputs, different mitigation strategies are employed. The majority of bias mitigation happens at inference time by modifying prompts~\cite{wan_survey_2024}. While this allows for correcting existing models, the approach is not fully controllable and may lead to model over-correction~\cite{wan_male_2024}. Furthermore, prompt modification can only work for known biases. Similarly, mitigation strategies that modify the model weights require prior knowledge of biases. The bias mitigation in DALL·E 2~\cite{nichol_dalle_nodate} balances the data in the training dataset to compensate for previously known biases. 

\subsection{Bias Exploration Systems}

Interactive VA systems for bias exploration are mostly studied in the context of textual representations, e.g., word embeddings. FAIRVIS~\cite{cabrera_fairvis_2019}, for example, uses multiple coordinated views to showcase model performance for different subgroups (e.g. races) in the outputs of a classifier. WordBias~\cite{ghai_wordbias_2021} uses word embeddings to visually explore intersectional bias by intersecting subgroups (e.g. \emph{``male''} and \emph{``Islam''}).

In contrast to text, in the field of image generation, bias exploration systems are either limited to pre-defined prompts~\cite{luccioni_stablediffusionbiasexplorer_2023} or have limited interactivity~\cite{dinca_openbias_2024, liu_organizing_2024, chinchure_tibet_2024}. Bias exploration systems, such as OpenBias~\cite{dinca_openbias_2024} or TIBET~\cite{chinchure_tibet_2024}, operate autonomously with an open set of biases created via LLM-generated prompts. The LLM-based biases (obtained by prompting the model to provide a list of potential biases), along with images generated from a prompt database, are fed into a VQA system. This system analyzes the frequency with which a proposed bias occurs in the images, resulting in the final bias assessment. While this approach allows for arbitrary biases to be inspected, it relies on an LLM (a non-user-controlled black box that itself may be biased) to have an understanding of the socio-cultural contexts from which bias may arise.

\rev{Beyond automated systems, interactive approaches have been proposed for model explainability. Human-in-the-loop interfaces for concept discovery~\cite{wang_drava_2023, zhao_human---loop_2022, huang_conceptexplainer_2022} focus on explaining the model behavior of large vision-language models, though their analyses are typically performed on static, non-generative datasets. 
VLSlice~\cite{slyman_vlslice_2023} offers an interactive interface for open-ended bias discovery in vision-language models by leveraging scatterplot-based visualizations to explore bias tendencies. However, it is limited in its flexibility as it only allows for limited human control over the bias dimensions. Next to the main prompt only a single initial baseline text is provided by the user. Furthermore, while it also utilizes a vision-language model for computing image-text similarities, there is no direct mechanism for users to investigate the trustworthiness of the model outputs.}

\subsection{User Interfaces for T2I Systems}

A core contribution of \name~is its novel prompting interface. To put it in context, we survey different approaches to interacting with T2I systems. The traditional text-to-image pipeline involves hand-written prompts provided by the user to the T2I system. This has led to research on how to effectively convey user intent to the image generator (i.e. prompt engineering)~\cite{liu_design_2022, mahdavi_goloujeh_is_2024, oppenlaender_taxonomy_2023}. A common approach to prompt engineering is to employ LLMs for prompt generation or prompt expansion, paired with a visualization of the generated images in some embedding space~\cite{feng_promptmagician_2023, brade_promptify_2023} or with direct manipulation of model attention for image refinement~\cite{wang_promptcharm_2024}. Alternatively, the prompting journey of the user is visualized as a graph that contains the previous output images in conjunction with weights showcasing the influence certain words have on a prompt~\cite{guo_prompthis_2024}. While the above-mentioned VA systems rely on (augmented) text input, some other approaches propose different input modalities. Almeda et al.~\cite{almeda_prompting_2024}, for example, employ a spreadsheet-based interface with commands for stylization or automatic prompt modification. 

Using graphs as an input modality has been explored for LLM prompting. Sensecape~\cite{suh_sensecape_2023}, an interface for sensemaking and exploration via LLMs, uses a graph-based layout to organize prompt variations for complex information tasks. Similarly, ChainForge~\cite{arawjo_chainforge_2024} provides users with a graphical, node-based interface for testing LLM robustness. In the context of T2I systems, however, graphs are so far only used as an intermediate data structure for scene organization via scene graphs~\cite{shukla_scene_2023, farshad_scenegenie_2023} and not as a direct input modality. To the best of our knowledge, the combination of an interactive graph-based prompting interface with bias exploration is a novel approach.

\section{\name}
\label{vibex}

We introduce \name~by first discussing \rev{design challenges} (Section~\ref{sec:requirements}) and the anticipated workflow (Section~\ref{sec:workflow}). Based \rev{on this, we describe our} bias probing approach (Section~\ref{sec:bias}), \rev{and} the application design (Section~\ref{sec:application}). 

\subsection{\rev{Design Challenges}}
\label{sec:requirements}

During exploratory visual analysis, users iteratively question the data, visually inspect the data, and refine their questions and hypotheses accordingly.  
Here, we list the most essential general \rev{characteristics} of exploratory visual analysis systems\rev{, as discussed by Battle and Heer~\cite{battle_characterizing_2019}, and elaborate on how they pose design challenges for} exploratory analysis of visual bias. 

\begin{description}
    \item[\textbf{\rev{C1:}}] \textbf{Flexibility. } Exploratory visual analysis systems need to support systematic and flexible queries so that all possible measures, dimensions, and features of the dataset can be studied by the user~\cite{perer_systematic_2008}. For visual bias exploration, this means that users should be able to flexibly test concepts that go beyond a pre-selected set of prompts.
    A special challenge, thereby, is that any user queries need to be correctly interpreted by the system.  
    \item[\textbf{\rev{C2:}}] \textbf{Real-time interactivity. } None of the previous works on bias in T2I systems report to operate in real-time, thereby not supporting the rapid pace of exploration needed in interactive visual exploration systems~\cite{battle_characterizing_2019}. \rev{The main challenge here is to develop a strategy for efficient evaluation of a large number (i.e. hundreds) of images in real-time.}  
    It has been shown that immediate approximate visual feedback is preferable over a slow and blocking visualization interface~\cite{zgraggen_how_2017}. Thus, it is essential to show at least an approximation of a response to the user's bias query without noticeable latency. 
    \item[\textbf{\rev{C3:}}] \textbf{Trustworthiness. } Understanding the correctness of the visualized data is crucial to exploratory VA systems~\cite{alspaugh_futzing_2019}. This notion of trust is especially important when designing applications for generative AI~\cite{weisz_design_2024}. Bias metrics are not always fully reliable since the bias evaluation system itself may be biased~\cite{wan_survey_2024, chinchure_tibet_2024} or because the sample based upon which the bias is computed is not representative~\cite{fabbrizzi_survey_2022}. In addition, a systematic visual difference observed for a test concept $t$ may not necessarily constitute a bias~\cite{dinca_openbias_2024, blodgett_language_2020}. In the end, it is up to the user to decide which concepts represent a bias~\cite{blodgett_language_2020}. Following a quick approximate response to a user's bias query ($\rightarrow$ \textbf{\rev{C2}}), the user, therefore, needs to be able to verify the system response through deeper inspection. 
    \item[\textbf{\rev{C4:}}] \textbf{Generalizability. }Any exploratory VA system should generalize beyond a single dataset. Similarly, \name~should be capable of analyzing images from different T2I models. 
    
\end{description}

\subsection{The \name~Workflow}
\label{sec:workflow}

\begin{figure}[t]
    \centering
    \includegraphics[width=\linewidth]{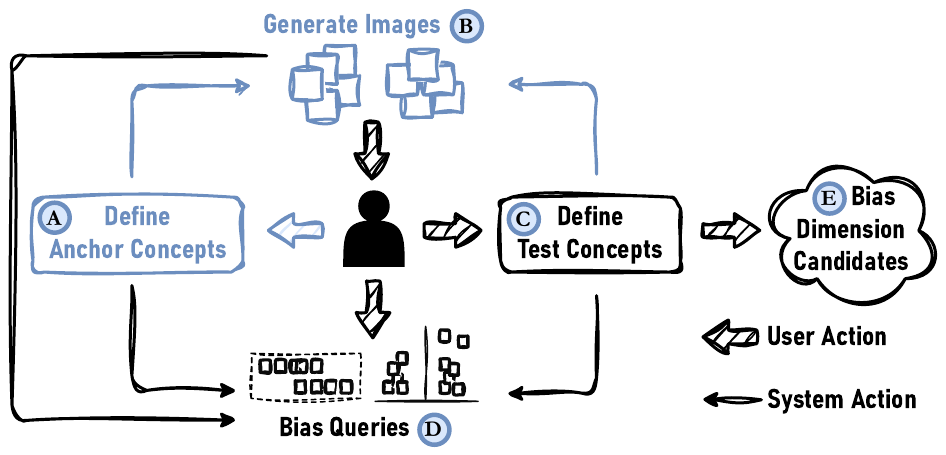}
    \caption{The \name~workflow: Users define anchor and test concepts and subsequently perform bias queries, resulting in bias candidates. \rev{Elements in blue represent a non-real-time operation.}}
    \label{fig:vibexWorkflow}
\end{figure}

The first step of the \name~workflow (Figure~\ref{fig:vibexWorkflow}) is to \blueletter{A} define two \textbf{anchor concepts} $c_1$ and $c_2$, with $c_j \in C$ (e.g., \emph{``woman''} and \emph{``man''}) between which to test for systematic visual differences. We generate $n$ sample images per anchor concept \blueletter{A} $\rightarrow$ \blueletter{B}. In practice, we chose $n=50$ because prior work used the same number of images~\cite{naik_social_2023} and our bias measures can be computed reasonably fast using this number (see Section~\ref{sec:zeroShotEvaluation} for performance measures). \rev{The image generation is a blocking operation that is currently not possible in real-time due to its high computational demands. However, all actions up to this point are performed only once per exploration session}. 

Users start their exploratory analysis by inspecting the generated images in the \textbf{data view} (Section~\ref{sec:dataView}) and iteratively adding, refining, or potentially removing \textbf{test concepts} $t \in T$ that describe visual properties of the image content in the \textbf{prompting tree} (Section~\ref{sec:promptingTree}) \blueletter{C}. For example, a test concept could state that the person depicted in the image is smiling. \name~automatically visualizes an approximate bias quantification based on \textbf{zero-shot bias probing} (Section~\ref{sec:bias}) \blueletter{C} $\rightarrow$ \blueletter{D}. Users perform in-depth confirmatory analysis by visually comparing the zero-shot similarity distributions between images representing the two anchor concepts and selected test concepts in different visualizations (Section~\ref{sec:biasVisualizations}) \blueletter{D}. 

Since our workflow builds upon image samples representing anchor concepts, there is a trade-off to consider: trustworthiness (\textbf{\rev{C3}}) vs.~real-time interactivity (\textbf{\rev{C2}}). The more images, the more reliable the bias quantification, but the longer they need to compute. In some cases, the test concept of interest may be underrepresented or not even depicted at all in any of the generated anchor concept images. For example, in Figure~\ref{fig:stripPlot}, the user tests for \emph{``black person''}, however, no persons with black skin color are present in the anchor concept images.
Thus, for testing T2I-models, \name~also supports inverse bias queries: instead of testing whether a test concept $t$ is biased towards one of the anchor concepts in $C$, we let users generate $m$ images representing a test concept (e.g., \emph{``An image of a smiling person''}) \blueletter{C} $\rightarrow$ \blueletter{B} and, finally, visualize, for each generated test image, the relative similarity to the anchor concepts \blueletter{B} $\rightarrow$ \blueletter{D}. 
Practically, the number of generated images $m$ representing the test concept will be considerably smaller than $n$ to reduce waiting times. For \name, we chose $m=5$. \rev{Based on the insights gained from the bias visualizations the user finally forms a selection of bias dimension candidates \blueletter{E} as a final output.}

\subsection{Zero-Shot Bias Probing}
\label{sec:bias}

\rev{Our approach to bias quantification is built upon measuring the compatibility} between a (generated) image $\mathcal{I}$ and a natural language text $\mathcal{T}$ representing a concept. \rev{To achieve this, \name~leverages the contrastively pre-trained model CLIP~\cite{radford_learning_2021}, which aligns visual and textual information in a shared embedding space.} 

Let $\mathbf{e_{\mathcal{I}}}$ and $\mathbf{e_{\mathcal{T}}}$ be two $d$-dimensional \rev{embeddings for an image} $\mathcal{I}$ and a text $\mathcal{T}$, respectively. We \rev{define} the similarity between \rev{these two embeddings} as a normalized dot product: 
\begin{equation}
    s(\mathcal{I}, \mathcal{T}) = 
    \frac{\mathbf{e_{\mathcal{I}}} \cdot \mathbf{e_{\mathcal{T}}} + 1}{2}.  
    \label{eq:similarity}
\end{equation}
A similarity value of $0$ \rev{indicates} no correlation, while $1$ \rev{denotes perfect alignment}.

\name~supports two types of bias queries. For \textbf{forward bias queries} \rev{(FBQs)}, we compute the similarity $s(\mathcal{I}_{jk}, t)$ between a textual test concept $t$ (e.g., \emph{``An image of a smiling person''}) and the $k$'th image associated with anchor concept $c_j$ (e.g., generated from the prompt \emph{``An image of a woman''}). For \textbf{inverse bias queries} \rev{(IBQs)}, we compute $s(\mathcal{I}_{tk}, c_j)$ for the $k$'th generated image representing the test concept $t$ (e.g.,  generated from the prompt \emph{``An image of a smiling person''}) against a natural language representation of anchor concept $c_j$ (e.g., \emph{``An image of a man''}). These similarity values serve as a foundation for our bias visualizations for in-depth confirmatory visual analysis based on images (Section~\ref{sec:biasVisualizations}).

To compute an approximate bias of a test concept $t$ towards an anchor concept $c_j$ based on forward bias queries, we use Bayes' theorem, which expresses the probability of the anchor concept $c_j$ to be true under the condition that 
the test concept $t$ is true: 
\begin{equation}
    P(c_j \mid t) = \frac{P(t \mid c_j) \cdot P(c_j)}{P(t)}. 
    \label{eq:bayes}
\end{equation}
Since we generate the same number of $n$ images per anchor concept, the prior \rev{is uniformly set to} $P(c_j) = 1 / |C|$. \rev{Therefore, with two anchor concepts $P(c_j) = 0.5$}. \rev{The likelihood} $P(t \mid c_j)$ is \rev{computed as the average similarity between test concept $t$ and all images generated for anchor concept $c_j$}: 
\begin{equation}
    P(t \mid c_j) = \sum_{k=1}^{n} \frac{s(\mathcal{I}_{jk}, t)}{n}. 
    \label{eq:distance_image_test_concept}
\end{equation}
\rev{The evidence} $P(t)$, finally, describes the \rev{overall} average similarity of the test concept $t$ \rev{across all $2 \cdot n$ generated images}. Intuitively, \rev{if the difference} $| P(c_1 \mid t) - P(c_2 \mid t) |$, \rev{is small, then the test concept $t$ does not strongly favor one anchor over the other, suggesting lower bias}. We use Equation~\ref{eq:bayes} to instantly show test concept biases directly in the prompting tree via color coding (Section~\ref{sec:promptingTree}).

Zero-shot bias probing tackles multiple \rev{design challenges}: 1) flexibility to formulate any test concept as a query to the set of anchor images (\textbf{\rev{C1}}) and 2) foundation to provide instant bias quantification based on a simple dot product computation (\textbf{\rev{C2}}).
Since CLIP can encode any input image, it is independent of the T2I model. Thus, zero-shot bias probing is also 3) a model-agnostic approach (\textbf{\rev{C4}}).

\subsection{\name~Application Design}
\label{sec:application}

To showcase the real-world applicability of \name, we built a reference implementation for a bias exploration interface in line with the \name~workflow. The web-based interface consists of multiple coordinated views (see Figure~\ref{fig:interface}): the data view, the bias visualizations, and the prompting tree.
Adding a new test concept automatically refreshes all visualizations and newly generated images instantly populate relevant views. Brushing and linking across all components enable quick filtering and highlighting. Additionally, users can switch between sessions with different data sources, supporting diverse scenarios across multiple T2I models (\textbf{\rev{C4}}).

\subsubsection{Data View}
\label{sec:dataView}

The data view (Figure~\ref{fig:interface} \redletter{1}) is a simple scrollable grid view, allowing the user to browse all generated images. Being able to view the dataset, the user may perform some free, unguided exploration here as a starting point for defining test concepts in the prompting tree. Interacting with the data view, the user will see the prompt from which an image was generated on hover. The images displayed in the data view are filtered or highlighted with a red outline based on interactions in other views. For instance, when the user hovers over a node in the prompting tree, all images generated from the node's prompt will be highlighted.

\begin{figure}[t]
    \centering
    \includegraphics[width=0.9\linewidth]{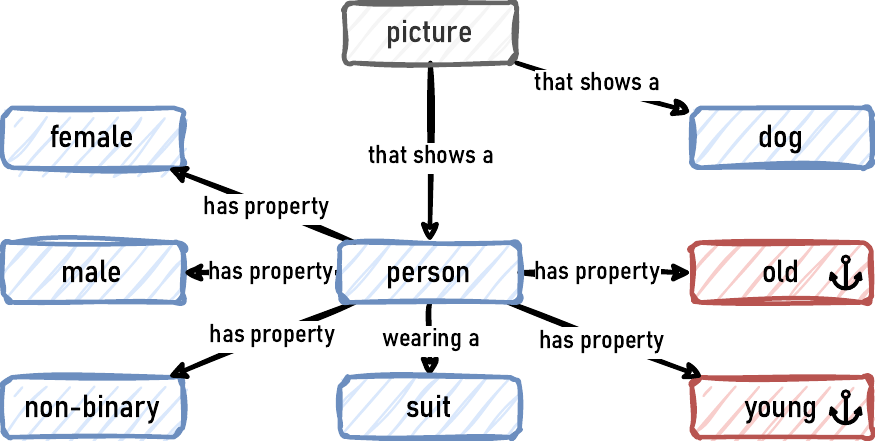}
    \caption{Schematic overview of the prompting tree. The root node \emph{``picture''} is part of all prompts. Anchor concepts (in red) represent the classes $c_1, c_2 \in C$, for which we probe different test concepts $t_i \in T$ (blue) for potential bias. The relation type between two concepts is indicated by the edge label. From this tree we may parse prompts such as \emph{``picture that shows a young person''} or \emph{``picture that shows a female person wearing a suit''}.}
    \label{fig:promptingTree}
\end{figure}

\subsubsection{Prompting Tree}
\label{sec:promptingTree}

The prompting tree (Figure~\ref{fig:interface} \redletter{3}) is the central user interface element of \name. It serves two main purposes: First, as input method, it supports flexible and systematic bias exploration by gradually adding and adjusting nodes that represent test concepts (\textbf{\rev{C1}}). It thereby helps users keep track of the bias candidates they have already investigated. Second, the prompting tree also serves as output method to provide instant zero-shot bias probing feedback (\textbf{\rev{C2}}).  

\begin{figure}[t]
    \centering
    \begin{subfigure}[]{0.48\linewidth}
        \centering
        \includegraphics[width=\textwidth]{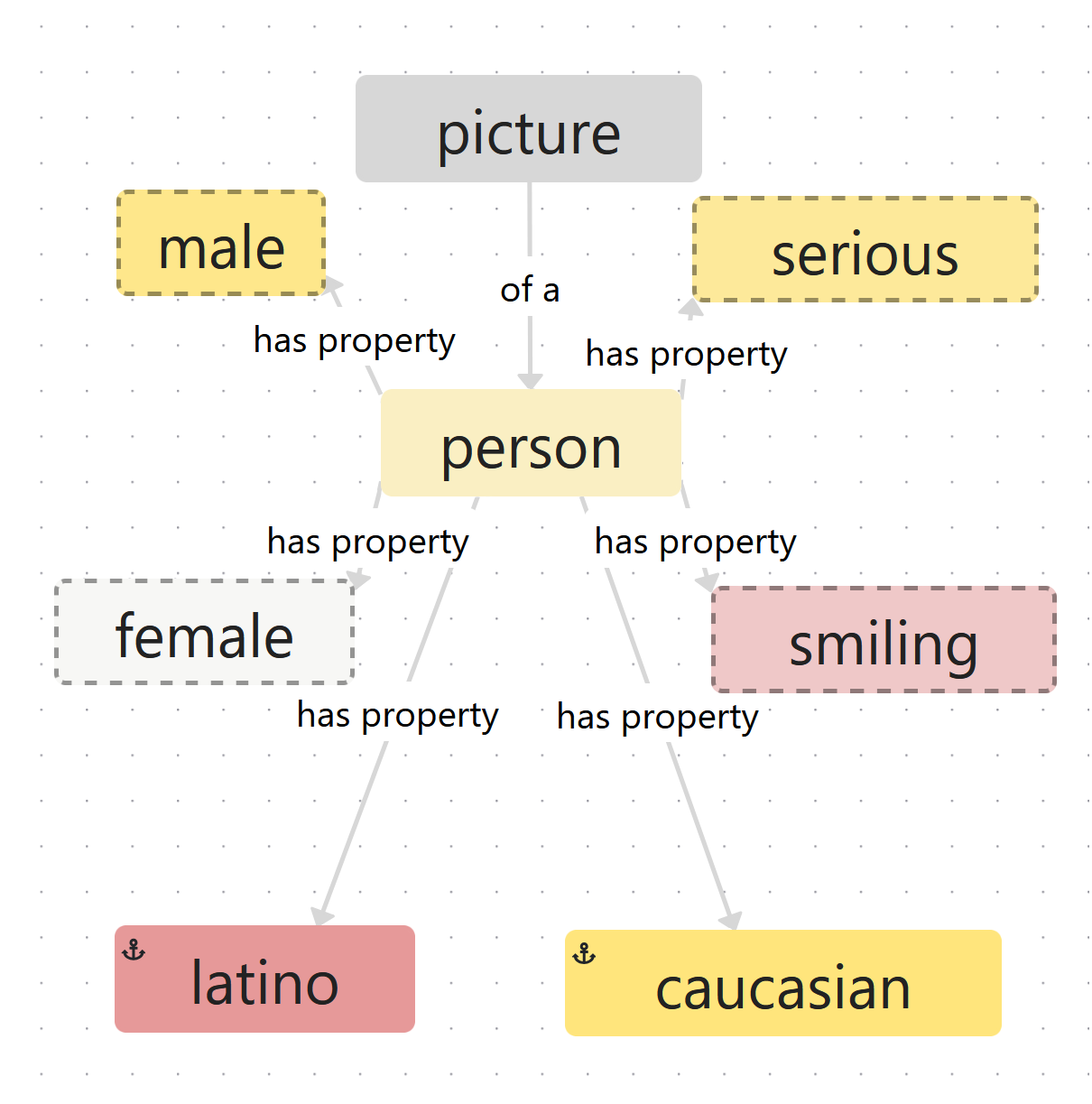}
        \caption{Generated Images (SD3)}
        \label{fig:promptingTreeSD3}
    \end{subfigure}
    \hfill
    \begin{subfigure}[]{0.48\linewidth}
        \centering
        \includegraphics[width=\textwidth]{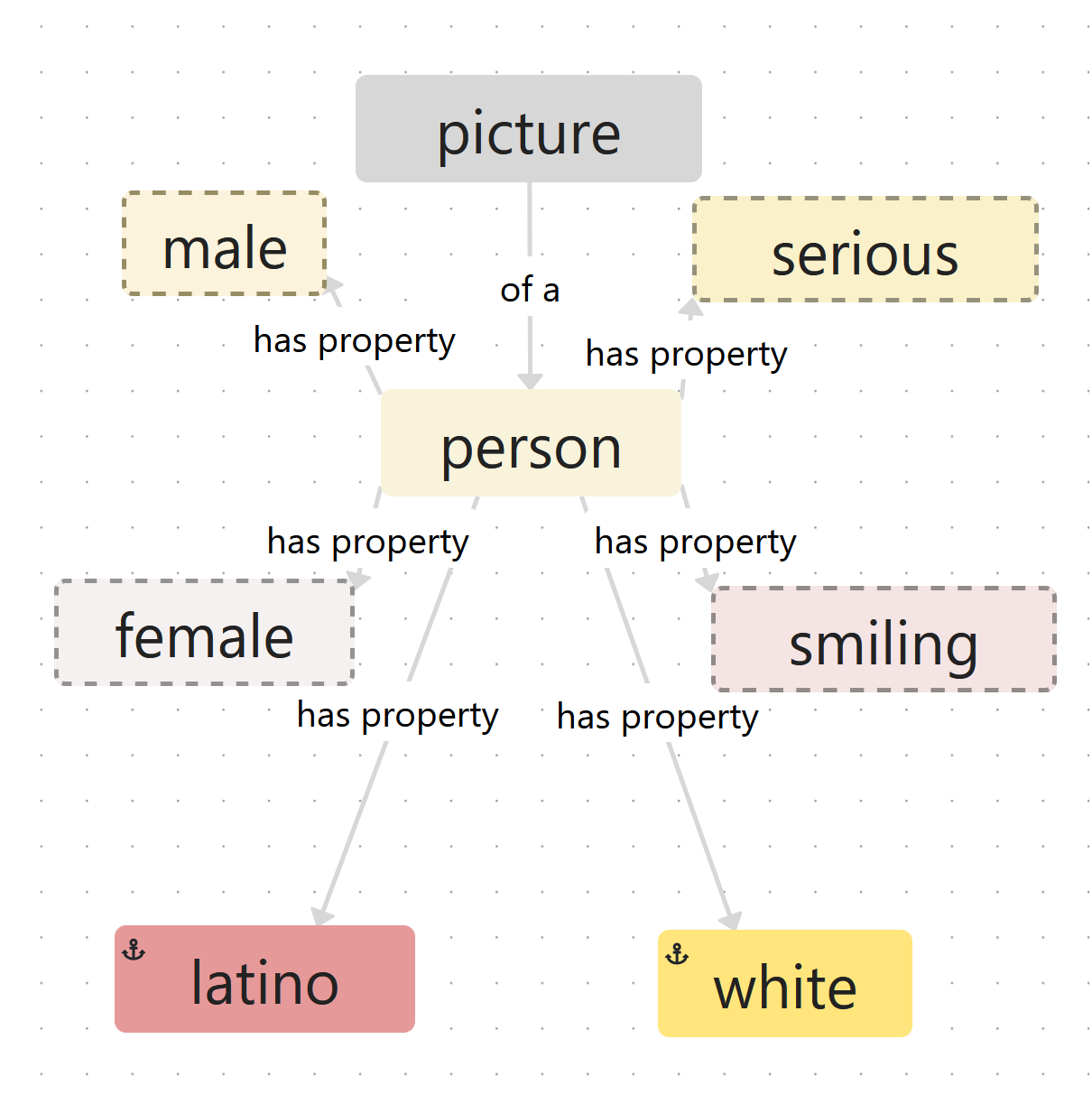}
        \caption{Images from FairFace}
        \label{fig:promptingTreeFF}
    \end{subfigure}
    \caption{Two prompting trees with the same concepts but differing data sources. The data loaded from the FairFace dataset is gender-balanced. Thus, gender is expected to be neutral. The minimal imbalance hints at a bias in CLIP. The SD3 data shows a pronounced bias toward \emph{``caucasian''} for the test concept \emph{``male''}.}
    \label{fig:promptingTreeExample}
\end{figure}

Formally, the prompting tree is a directed acyclic graph $G=(V, E)$ where $V = \{v_0\} \cup \{v \mid v \in C \text{ or } v \in T\}$ are $k$ concepts representing the anchor concepts $C$ and a variable number of test concepts $T$, with $v_0$ being the root node. $E = \{ (v_i,r,v_j) \mid v_i,v_j \in V, r \in R\}$ are labeled edges between two nodes with labels contained in the relation set $R$. The relation set $R$ contains the connecting words, which allow for parsing the tree structure into natural language. For the example tree in Figure~\ref{fig:promptingTree} the relation set is defined as $R = \{\text{\emph{``has property''}},\text{\emph{``that shows a''}},\text{\emph{``wearing a''}}\}$. The \emph{``has property''} relation is the default edge label, which is used to attach adjectives to concepts.  
Anchor icons within the nodes indicate anchor concepts, and image icons indicate the test concepts from which images were generated by the user for inverse bias queries. Additionally, as the user may define test concepts for more in-depth bias probing (see Section~\ref{sec:biasVisualizations}), the respective nodes will be marked with dashed outlines (see Figure~\ref{fig:promptingTreeExample}).

Every concept expressed as a node in the tree can be parsed into a natural language representation $\mathcal{T}$ that serves as input to our similarity computation (Equation~\ref{eq:similarity}), as well as potential prompt to an image generator for inverse bias queries. 
A text representation of a concept $\mathcal{T}$ is constructed by concatenating node and edge labels along a branch of the tree, starting at the root node. For example, in Figure~\ref{fig:promptingTree}, the branch ending with the concept \emph{``female''} would be parsed in the following way: $(\text{``picture''} \oplus \text{``that shows a''} \oplus \text{``female''} \oplus \text{``person''})$, where $\oplus$ denotes a concatenation.
If multiple branches are selected, they will be parsed into a combined text. If one node has multiple \emph{``has property''} relations, the corresponding labels are chained together (e.g., \emph{``young male person''}), while all other relations at the same level are connected with ``and'' (e.g., \emph{``picture that shows a person and a dog''}).
With this definition, nodes generally represent nouns and adjectives, while the edges are verbs and prepositions.
Thus, the prompting tree adheres to design guidelines for T2I prompting, which suggest that effective prompting should ``\emph{focus on subject and style keywords over connecting words}''\cite{liu_design_2022}. 
\rev{While a list of concepts could encode the same information as the prompting tree, the tree-based representation avoids redundancies by allowing for concept reuse: a comprehensive list would contain $|V|$ prompts with the root node being redundantly represented in each sentence. In the case of inverse bias queries, it is possible to combine multiple branches of the tree into a single prompt, leading to a combinatorial explosion of possible prompts. Furthermore, due to the ability to expand the tree at any node, it is possible to ``flip'' a concept (e.g. adding \emph{``no''} to the concept \emph{``hair''} using the relation \emph{``has property''} to indicate baldness). A potential alternative representation would be a Word Tree~\cite{wattenberg_word_2008}, although here, the more constrained structure would hinder manual re-ordering and custom spatial organization through, for example, clustering of similar concepts by the user.}

To visually encode the result of the zero-shot bias probing, we color-code the test concept nodes according to the computed probabilities from Equation~\ref{eq:bayes}. We use a diverging color scale, where the endpoints represent the colors associated with the two anchor concepts. The anchor concept colors can be selected by the user from a pre-defined list of colors. Gray represents an unbiased test concept, where $P(c_1 \mid t) \sim P(c_2 \mid t) \sim 0.5$. If an anchor concept's probability nears 1, the test concept's color will closely match the anchor concept's color. With that, the user can very quickly grasp which test concepts are similar to an anchor concept according to zero-shot bias probing. For example, Figure~\ref{fig:promptingTreeSD3} shows \emph{``serious''} as more similar to \emph{``caucasian''}, while \emph{``smiling''} is closer to \emph{``latino''}.

\begin{figure}[t] 
\centering 
\includegraphics[width=\linewidth]{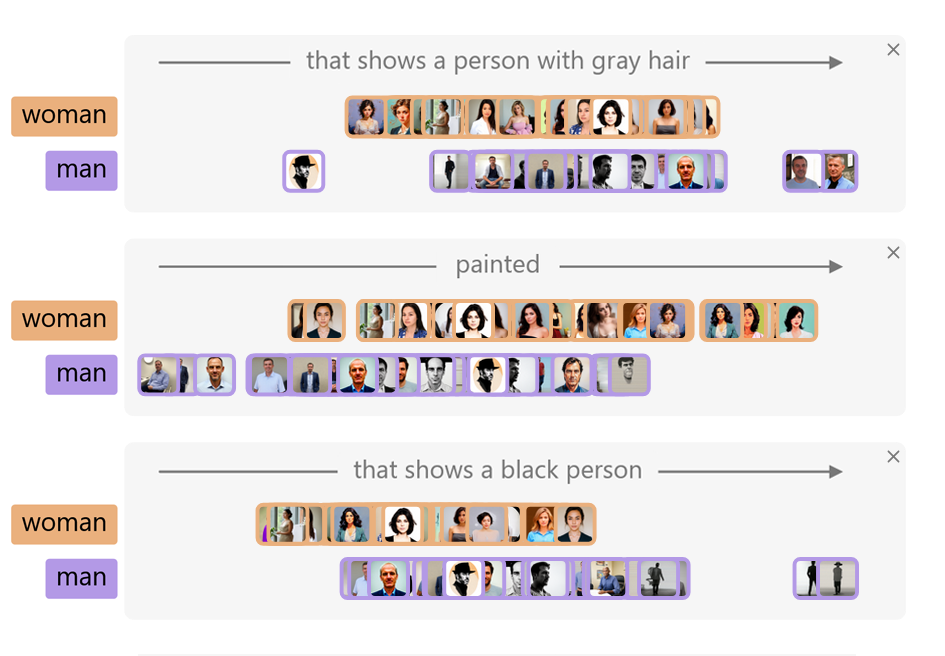} 
\caption{Strip plots for three test concepts (\emph{``gray hair''}, \emph{``painted picture''}, \emph{``black person''}). The univariate distribution of image text similarities is plotted for each anchor concept $c_i \in C$, here \emph{``woman''} and \emph{``man''}. Note how the highest scoring image for \emph{``black person''} is a silhouette picture.} 
\label{fig:stripPlot} 
\end{figure} 

\subsubsection{Bias Visualizations}
\label{sec:biasVisualizations}

The advantage of zero-shot bias probing based on CLIP is that it provides instant bias estimates. The downside is that these estimates are not always reliable because CLIP itself is biased~\cite{hamidieh2024identifying,10.1145/3442188.3445932}. This is illustrated in Figure~\ref{fig:promptingTreeExample}: using the prompting tree, we discovered a bias of the test concept \emph{``male''} towards the anchor concept \emph{``caucasian''}. We tested this discovery against a hand-picked gender- and race-balanced sample from FairFace~\cite{karkkainen_fairface_2021}, where images of male persons are evenly distributed across images labeled as \emph{``latino''} and \emph{``white''}. Our test concept bias metric (Equation~\ref{eq:bayes}) also indicates a weak bias towards images labeled as \emph{``white''} when testing against \emph{``picture of a male person''}, although this bias is not present in the anchor images. This could be caused by CLIP interpreting images of Caucasian persons to be ``more male'' than images of Hispanic people. Another problem is that text representations of concepts may be interpreted differently by the vision-language model than intended by the user. For example, in Figure~\ref{fig:stripPlot}, the test concept \emph{``picture that shows a black person''} has very strong similarities with images showing a dark silhouette, while people with dark skin are not even present in the anchor concept image set. Similarly, we could not use the FairFace label \emph{``white''} to generate the anchor images for the example in Figure~\ref{fig:promptingTreeExample}: SD3 would generate images of people with their faces painted white, so we used \emph{``caucasian''} instead. 

To allow for real-time interactivity (\textbf{\rev{C2}}) while maintaining trustworthiness (\textbf{\rev{C3}}), it is therefore essential to keep the human in the loop. We support a human-centered bias exploration loop through multiple visualizations (Figure~\ref{fig:interface} \redletter{2}) offering in-depth confirmation of observed or suspected biases. For in-depth bias probing, users select one or multiple nodes in the prompting tree to generate a prompt of a test concept to be investigated in more detail. 

For in-depth visual inspection of forward bias queries, we provide juxtaposed \textbf{strip plots}, where the $x$ axis corresponds to the similarity distribution of all anchor concept images towards the selected test concept (computed according to Equation~\ref{eq:similarity}), separated by anchor concept on the $y$ axis. The strip plots in Figure~\ref{fig:stripPlot} confirm that only anchor images for \emph{``man''} depict persons with gray hair and that primarily images associated with \emph{``woman''} are painted.

\begin{figure}[t]
    \centering
    \includegraphics[width=0.7\linewidth]{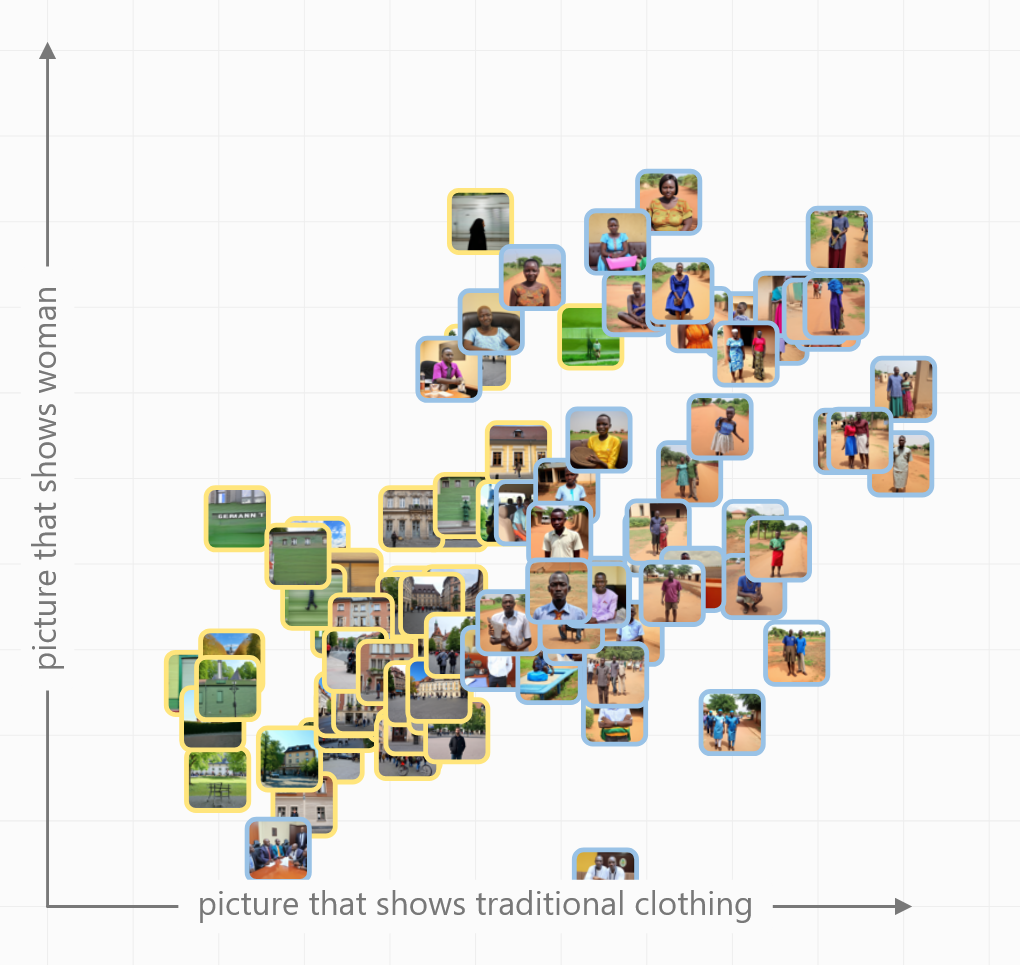}
    \caption{Intersectional bias plot for \emph{``woman''} and \emph{``traditional clothing''} with the anchor concepts \emph{``Germany''} and \emph{``Nigeria''}. Images in the top right are Nigerian women in traditional clothing, while for Germany no traditional clothing is present.}
    \label{fig:intersectionalBias}
\end{figure}

Users may also visualize bivariate distributions of forward bias queries for two test concepts using the \textbf{intersectional bias scatterplot}. 
All images generated from the anchor concepts $C$ are placed at the $(x,y)$ coordinate corresponding to their similarities to the two selected test concepts. Each image is outlined with the color of its associated anchor concept. The resulting distribution allows the user to observe possible correlations between the two selected test concepts with respect to the anchor concepts. For example, Figure~\ref{fig:intersectionalBias} illustrates an intersection discovered by an expert in our case study between \emph{``woman''} and \emph{``traditional clothing''}, which only applies to the anchor concept \emph{``Nigeria''}.

\begin{figure}[t]
    \centering
    \includegraphics[width=0.98\linewidth]{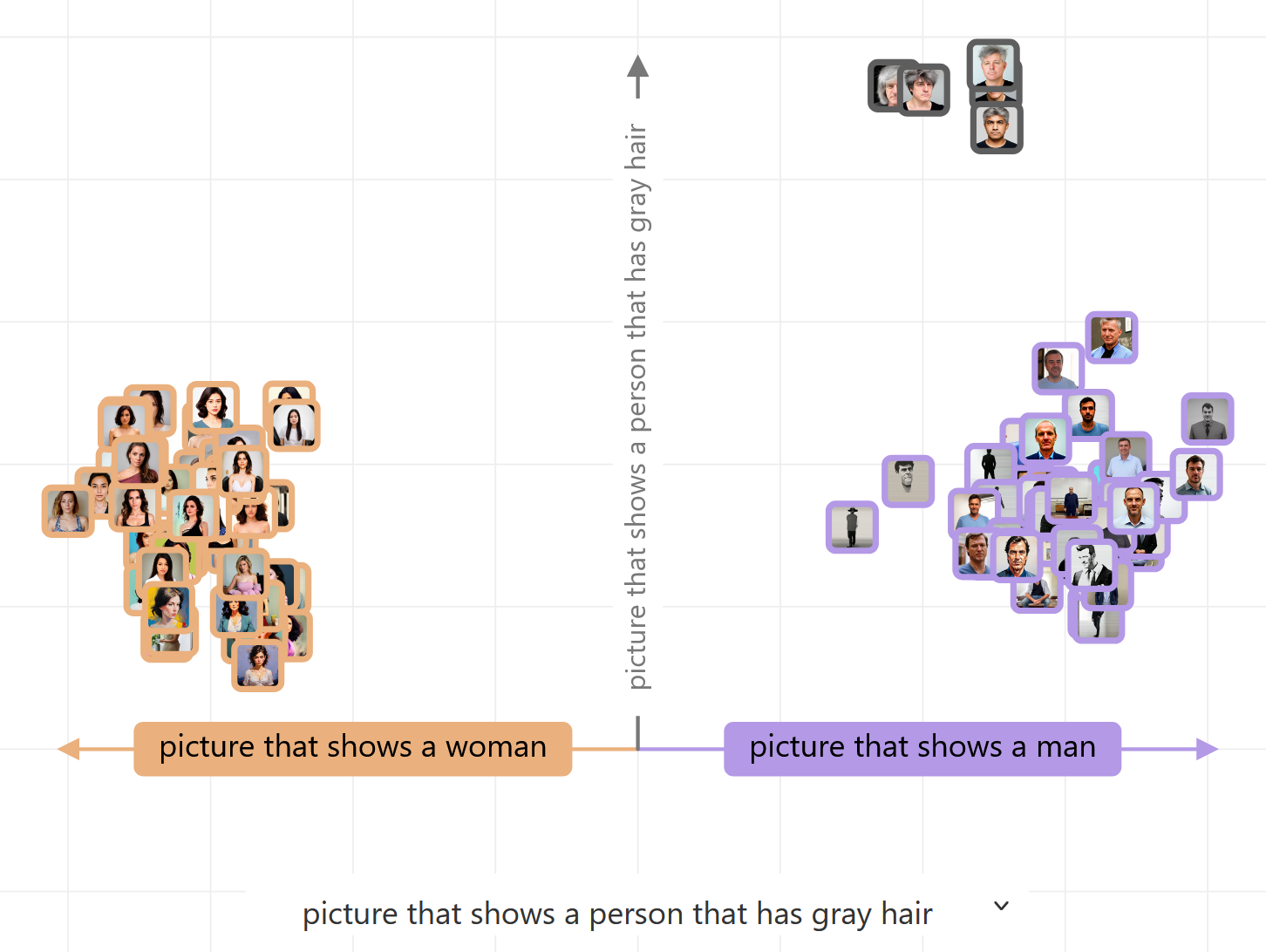}
    \caption{Inverse bias query plot for \emph{``picture that shows a person with gray hair''} for the anchor concepts \emph{``woman''} and \emph{``man''}. Images generated for the test concept have a gray outline. We see that \emph{``gray hair''} is predominantly associated with \emph{``man''}.}
    \label{fig:inverseQuery}
\end{figure}

Finally, results of inverse bias queries can be inspected in the \textbf{inverse bias scatterplot}. In this scatterplot, we show all anchor concept images, as well as images generated for the selected test concept (e.g., \emph{``picture of a person with gray hair''}). Anchor concept images are framed by the color associated with their respective anchor concept. Test concept images have a gray frame. The $x$ position of each image $\mathcal{I}_i$ represents the relative similarity to the textual representation of both anchor concepts: 
\begin{equation}
    x(\mathcal{I}_i) = s(\mathcal{I}_i, c_2) - s(\mathcal{I}_i, c_1). 
    \label{eq:x}
\end{equation}
We draw a vertical line at $x=0$ to visually separate the two anchor concepts. The $y$ axis represents the forward bias query towards the selected test concept (i.e., $s(\mathcal{I}_i, t)$). 
It can be expected that the anchor concept images are clearly separated horizontally, while the test concept images are separated from the anchor concept images vertically. This pattern is clearly visible in Figure~\ref{fig:inverseQuery}, where all anchor concept images for \emph{``woman''} (orange) are on the left, and all images for \emph{``man''} (purple) are on the right. Also test concept images showing \emph{``a person with gray hair''} (gray) are clearly more similar to their corresponding text representation and therefore shown on top. Most interestingly, all test concept images are clearly more similar to the anchor concept \emph{``man''} than to \emph{``woman''}, thereby confirming the observed forward bias in Figure~\ref{fig:stripPlot} that gray hair is more strongly associated with \emph{``man''}.

\section{Implementation}
\label{sec:implementation}

The focus of our prototype implementation of \name~is to \rev{address the design challenge} of real-time interactivity (\textbf{\rev{C2}}), while supporting the user interaction discussed in Section~\ref{sec:application}. \name~is implemented as a client-server application. The back-end is a Python server, which handles session management and computes image-text similarities according to Equation~\ref{eq:similarity}. Whenever the front-end reports changes to the prompting tree, the server computes and stores the similarity values between the text representation of the affected prompting tree node and all anchor images. While the server is running, we keep the CLIP model in memory and also cache the image embedding vectors, such that for a new similarity calculation, we only need to generate the embedding of the test concept and compute the similarity measure. When the new similarity values have been computed, the array of similarity values is synced with the front-end, which then updates the forward bias values for the affected nodes in the prompting tree using Equation~\ref{eq:bayes}. Upon image generation (which is performed on a dedicated server), an inverse bias query is triggered, updating the inverse bias plot. Since image generation can take up to a few minutes, the front-end shows $m$ placeholder images with a progress indicator in the data view until it gets notified that image generation is complete. 

The \name~front-end is a web application using the React Framework~\cite{react}, with D3.js~\cite{d3js} for the bias visualizations while the prompting tree is based on React Flow~\cite{reactflow}. The back-end Python server uses CLIP \texttt{laion/CLIP-ViT-bigG-14-laion2B-39B-b160k}~\cite{clip_vit_laion}. With the system being model-agnostic, we can utilize different image generation APIs either from local models (e.g., SD3) or from repositories such as \emph{Hugging Face}.

\section{Evaluation}
\label{sec:evaluation}

We performed multiple evaluations to test \name~against our \rev{design challenges} formulated in Section~\ref{sec:requirements}. To test for flexibility (\textbf{\rev{C1}}), we performed a pilot study to test how confidently and correctly users can express observed and expected biases in the prompting tree (Section~\ref{sec:promptingTreeStudy}). To evaluate real-time interactivity (\textbf{\rev{C2}}), we measured the computation times of zero-shot bias queries using CLIP (Section~\ref{sec:zeroShotEvaluation}). Trustworthiness (\textbf{\rev{C3}}) was evaluated through a combination of CLIP-based similarity results compared to a ground truth (Section~\ref{sec:zeroShotEvaluation}) and an expert case study, where the discovered biases were a-posteriori validated using traditional non-interactive bias measures (Section~\ref{sec:expertStudy}). Finally, we cross-validated confirmed biases with prior work \rev{(including an automated method)} to filter out which visual biases experts could discover using \name~that have not yet been reported in literature. 

For all evaluations, we utilize SD3 as a state-of-the-art diffusion model. SD3 does not employ explicit bias mitigation techniques, such as prompt injection, allowing for direct evaluation of its intrinsic bias. All evaluations are based on three selected anchor concept pairs, which were already investigated in prior work. 
Our scenarios were selected based on prior work by Naik and Nushi~\cite{naik_social_2023}. In their paper, they investigated the anchor concept dimensions \emph{gender, age, race}, and \emph{geographical location} with respect to neutral test concepts, which were different occupations, personality traits, and everyday situations. 
We selected the scenarios \emph{gender} (S\textsubscript{gender}), \emph{race} (S\textsubscript{race}), and \emph{geographical location} (S\textsubscript{loc}) and derived corresponding anchor concepts listed in Table~\ref{tab:study-scenarios}. In addition, we created a scenario S\textsubscript{train} for the training session in our two user studies (Section~\ref{sec:promptingTreeStudy} and Section~\ref{sec:expertStudy}). The corresponding anchor concept images can be found in Appendix~\ref{app:biasScenarios}.

\begin{table}[ht]
\centering
\small
\begin{tabular}{lll}
\hline
\textbf{Scen.}   & \textbf{$\mathcal{T}(c_1)$}              & \textbf{$\mathcal{T}(c_2)$} \\ \hline
S\textsubscript{train} & drawing of a old-timer car        & drawing of a futuristic car  \\ 
S\textsubscript{gender}     & picture that shows a woman      & picture that shows a man \\ 
S\textsubscript{race}       & picture of a Latino person     & picture of a Caucasian person  \\ 
S\textsubscript{loc}       & picture taken in Germany        & picture taken in Nigeria  \\  \hline
\end{tabular}
\caption{Anchor concept prompts used as evaluation scenarios.}
\label{tab:study-scenarios}
\end{table}

For these scenarios, using Stable Diffusion v1 and DALL·E as image generators, Naik and Nushi~\cite{naik_social_2023} reported a clear bias towards images being situated in Western countries and depictions of poor economic conditions being associated with Nigeria (S\textsubscript{loc}). For the neutral prompt \emph{``person''} they furthermore reported a general bias towards white people and a model-specific gender imbalance.

\subsection{Prompting Tree Pilot Study}
\label{sec:promptingTreeStudy}

We conducted a pilot study to evaluate the flexibility and reliability of our prompting tree as \emph{input} method (\textbf{\rev{C1}}) in an early phase of the design process. The goal of the study was to assess whether users can flexibly express test concepts using the prompting tree and whether the test concepts can be parsed into a useful text representation following the procedure described in Section~\ref{sec:promptingTree}. 

Nine volunteer students and employees from a local university participated in the study (five females and four males, aged 24-36, all with a computer science background). No participant was familiar with the planned \name~application. Participants were equally distributed across the three scenarios listed in Table~\ref{tab:study-scenarios}, resulting in three participants per scenario. For each anchor concept, we generated $n=50$ images. The users' task was to describe all observed or suspected biases as labeled nodes in the prompting tree. Prior to the main task, we conducted a training task using scenario S\textsubscript{train} to explain how to add and remove nodes and relations, as well as how to define new relation types. For this study, we only showed the data view and used the prompting tree as the sole input method without visual encoding of zero-shot bias probing. 

After the study, we automatically parsed each node in the created prompting trees into a text representation. Two independent coders then categorized each parsed text into three categories: \emph{correct} (the text is meaningful), \emph{concept problem} (e.g., a typo or a missing word in one or more nodes), and \emph{relation problem} (e.g., an illogical connecting word or a grammatical error in one or more relations).

\paragraph*{Results:} 
A total of 121 nodes were created during the study, with seven to 22 per user (13 on average). The coders found 44 problems in total, but the majority of problems were contributed by two outlier users. 
One user had a 100\% error rate as they built a hierarchical graph that could not directly be translated into natural language. This resulted in parsed texts like \emph{``picture with classic architecture focus content''}. The other user did not adjust the relation qualifiers, keeping ``has property'' for all relations. This resulted in parsed texts like \emph{``picture of a formal wear person''}. Excluding these outliers, 78.7\% of texts were classified as \emph{``correct''},  7.8\% as \emph{``concept problem''} and 13.5\% as \emph{``relation problem''}. Relation problems were primarily caused by default \emph{``has property''} relations that were not adjusted appropriately. 

We compared a selected set of texts with relation problems to a corrected version with respect to CLIP-based similarity scores and resulting images when used as prompts for SD3 (see Appendix~\ref{app:clip-robustness}). It can be observed that both models have a certain level of resilience to grammatical errors, which confirms prior work that subject and style keywords are more important than connecting words~\cite{liu_design_2022}. 

The test concepts expressed in the prompting trees primarily contained rather obvious entanglements, such as a tendency towards \emph{``dark skin''} for the test concept \emph{``Nigeria''} in scenario S\textsubscript{loc}, towards
\emph{``man''} for \emph{``beard''}, and towards \emph{``woman''} for \emph{``long hair''} (S\textsubscript{gender}). More interestingly, all three participants inspecting scenario S\textsubscript{race} reported a gender bias, which was also previously described by Naik and Nushi~\cite{naik_social_2023}. Three users also reported systematic differences for test concepts related to clothing style in scenarios S\textsubscript{loc} and S\textsubscript{race}, respectively. This has also been reported from previous bias studies~\cite{naik_social_2023, liu_organizing_2024, dinca_openbias_2024}. 

When asked about special difficulties in the post-experiment interview, participants noted that the cognitive load required to parse and verify the text representation associated with a prompting tree node was challenging.
Overall, however, all participants considered the prompting tree to be a very engaging interface. 

We conclude that it is challenging to use the prompting tree with complete accuracy without training. However, commonly observed relation problems seem to have little influence on the model performance. To reduce the cognitive load for the user, we now display the fully parsed text when hovering over a node. This way, users can easily check and potentially correct their prompting tree.  

\subsection{Zero-Shot Bias Probing Performance}
\label{sec:zeroShotEvaluation}

\rev{We performed a limited evaluation of CLIP testing its ability to detect biases as well as its time performance. The details of this evaluation can be found in Appendix~\ref{app:clip-experiments}. In summary, we found that CLIP reliably detects an expected entanglement, while it shows no significant bias for a balanced image characteristic. This indicates that our zero-shot bias probing approach can provide sufficiently reliable results for a first approximate bias estimation (\textbf{\rev{C3}}).} We conducted the experiments using an NVIDIA H100 GPU with 80GB of memory. 

System logging revealed that a single CLIP-based similarity measure \rev{between one concept and 100 anchor images (with cached embedding vectors) requires $1.47\pm0.18$ seconds to compute}. This does not entirely \rev{satisfy our design challenge of} real-time \rev{interactivity (\textbf{C2})} since the impression of instant feedback usually requires response times below 0.1 seconds~\cite{nielsen1994usability}.
\rev{However, since we compute similarities on the server (see Section~\ref{sec:implementation}), these computations are not blocking}, and visual feedback is automatically updated one to two seconds after the user has modified the prompting tree.

\subsection{Expert Case Study}
\label{sec:expertStudy}

Finally, we evaluated whether expert users were able to discover visual bias dimensions with the support of \name~that were so far unknown to them. 
We invited four domain experts in the fields of AI, media ethics, and digital humanism to perform a bias exploration case study using the \name~interface. Users could select one or multiple scenarios to investigate from Table~\ref{tab:study-scenarios}. Three experts (E1, E2, and E4) explored S\textsubscript{gender}, E3 and E4 also examined S\textsubscript{loc}. We first performed a pre-experiment interview about their experience with generative AI after which the participants \rev{performed} a training task as described in Section~\ref{sec:promptingTreeStudy}. For all scenarios, we prepared $n=50$ generated images per anchor concept and a prompting tree containing the two anchor concepts. We then asked the experts to extend the prompting tree with test concepts based on what they observed in the anchor concept images and based on their expectations, respectively. The participants were encouraged to verbally express their actions by thinking aloud, which was recorded and transcribed for qualitative evaluation. After the study, we asked participants to list all test concepts that they considered to represent a bias. The individual sessions lasted between 60 and 90 minutes.

\subsubsection{Results}
\begin{table}[t]
\centering
\small
\begin{tabular}{lllll}
\hline
\textbf{Scenario} & \textbf{Test Concept}  & \textbf{Query} & \textbf{Tendency} & \textbf{Expert}  \\ \hline
\multirow{20}{*}{S\textsubscript{gender}}   & \textbf{person} & IBQ       & men      & E1          \\
                          & \textbf{beautiful}             & IBQ       & women    & E1, E4      \\
                          & naked shoulders        & FBQ       & women    & E1, E4       \\
                          & bare skin              & FBQ       & women    & E2         \\
                          & \textbf{long hair}              & FBQ       & women    & E1, E4      \\
                          & \textbf{dark skin}              & IBQ       & women    & E1          \\
                          & asian*            & FBQ       & women    & E2          \\
                          & \textbf{happy}            & IBQ       & men      & E1          \\
                          & \textbf{standing}               & FBQ       & men      & E1, (E4)     \\
                          & \textbf{doctor}*                 & FBQ       & men      & E2          \\
                          & \textbf{nurse}*                  & FBQ       & women    & E2          \\
                          & \textbf{black and white }       & FBQ       & men      & E4, (E2)     \\
                          & \textbf{bright colors}          & FBQ       & women    & E2          \\
                          & \textbf{old}*                    & FBQ       & men      & E2          \\
                          & young*                  & FBQ       & women    & E2          \\
                          & \textbf{business look}*          & FBQ       & men      & E4          \\
                          & \textbf{professional}*           & FBQ       & men      & E2          \\ 
                          & \textbf{boss}*                   & FBQ       & men      & E2          \\ 
                          & serious*                & FBQ       & balanced & E2          \\ 
                          & child*                  & FBQ       & balanced & E2          \\ \hline
\multirow{4}{*}{S\textsubscript{loc}} & traditional clothing*   & FBQ       & Nigeria  & E3, E4      \\
                          & sand-colored tones     & FBQ       & Nigeria  & E4          \\
                          & greenery               & FBQ       & Germany  & E4          \\
                          & classic architecture*   & FBQ       & Germany  & E4          \\ \hline 
\end{tabular}
\caption{Biases discovered via \name~during the expert studies. FBQ and IBQ show whether the concept was tested through a forward or inverse bias query; the tendency indicates towards which anchor concept the bias is expressed. \rev{The * marker signifies concepts that are covered by the bias axes automatically determined by TIBET~\cite{chinchure_tibet_2024}, while the tendencies for \textbf{bold} concepts were confirmed by our subsequent analysis (only performed for S\textsubscript{gender}).}}
\label{tab:discoveredBiases}
\end{table}

The test concepts for the discovered biases are summarized in Table~\ref{tab:discoveredBiases}, while all prompting trees are displayed in Appendix~\ref{app:expertTrees}. The three experts who explored S\textsubscript{gender} found 20 test concepts with potential bias. The majority (16) of these concepts were forward bias queries confirmed through a strip plot. E1 and E4 also performed inverse bias queries (for E2 there was a problem with the image generation server, so no inverse bias queries were conducted). Notably, especially through the inverse queries, some unexpected biases were found: SD3 has a general gender bias towards \emph{``male''} when prompted to generate a \emph{``picture that shows a person''}. Furthermore, \emph{``happy''} is skewed towards \emph{``man''} and both, \emph{``beautiful''} as well as \emph{``dark skin''}, toward \emph{``woman''}. All four experts used the intersectional bias plot to check for expected correlations. 

E3 found a potential intersectional bias in S\textsubscript{loc}, where Nigerian women appeared to be more frequently shown in traditional clothing (see Figure~\ref{fig:intersectionalBias}). The same tendency was also discovered by E4 in a forward bias query. 
Besides the bias candidates discovered through interaction with the interface, general observations of the images resulted in the conclusion that the \emph{``man''} images showed more diverse body types and framing in S\textsubscript{gender} (E4), while E1 noticed through their \emph{``person''} query that this ungendered prompt resulted in melancholic depictions of mostly light-skinned men.

All four experts noted that the tool was engaging and made them more aware of the problem biases in T2I pose. They mentioned that, although some of the tendencies were visible even without bias probing, it was the interactivity and instant feedback of the prompting tree that kept them engaged. The prompting tree still posed a challenge, with some grammatically incorrect relations (e.g., Appendix~\ref{app:expertTrees} Figure~\ref{fig:s2e02}). However, as discussed in Section~\ref{sec:zeroShotEvaluation}, the system proved to be robust to malformed text input. Apart from using the tool in their own research, experts also suggested that it would be well-suited for journalists and decision-makers to learn about biases in T2I models. Furthermore, they argued for \name~to be used as a basis for model auditing.

\subsubsection{Validation of Discovered Biases}
\label{sec:validation}

To verify that the biases discovered by experts constitute actual imbalances, we utilize the FairFace classifier~\cite{karkkainen_fairface_2021} as a trusted evaluation entity because it is trained on a balanced dataset. We pick those 20 biases from the case study that align with the FairFace classes (age, gender, and race), generate 50 new images per bias (1000 in total), and classify them. We generate new images to also compensate for a potential sample bias in the case studies. The images generated for our validation step are available on \href{https://osf.io/yd7gr/?view_only=08fa3822c66a4838a77b03cee4a7b8a6}{osf.io}.

\rev{We show the} results of our validation step \rev{in Table~\ref{tab:discoveredBiases}}, where we \rev{highlight all confirmed biases}. The FairFace classifier only operates on images with visible faces, thus we only classify such images. All four inverse bias queries (\emph{``person''}, \emph{``beautiful''}, \emph{``dark skin''}, \emph{``happy''}) are confirmed by the FairFace validation, indicating that the small sample of five images provides reliable insight. Furthermore, the expected gender bias for \emph{``nurse''}/\emph{``doctor''} by E2 is confirmed. For the forward bias queries with test concepts tending toward \emph{``woman''}, most tendencies are not confirmed, as they instead showcase a bias toward depicting men. This is probably caused by a general gender bias of SD3 towards men, which can be observed in Appendix~\ref{app:discovered-biases} Figure~\ref{fig:FFEval}. \rev{The test concept \emph{``young''} may be too ambiguous: while the experts used it to label young adults, SD3 produced images of kids. The concept \emph{``naked shoulders''}, on the other hand, may be too specific and, therefore, not well-represented in the training data. We discuss a potential strategy to alleviate the problem of these ``false positives'' in Section~\ref{sec:discussion}.
Only for }\emph{``long hair''}, \emph{``dark skin''} and \emph{``bright colors''}, the bias toward \emph{``woman''} is also clearly present in the validation. While \emph{``long hair''} was also reported in the pilot study based on image observations only, \emph{``dark skin''} and \emph{``bright colors''} were only discovered by the experts through the \name~workflow. To the best of our knowledge, these biases have not been reported by prior work. 

\rev{To further test whether an automated bias discovery system could find these bias dimensions, we queried TIBET~\cite{chinchure_tibet_2024} with the image generation prompts from S\textsubscript{gender} and S\textsubscript{loc}. We performed the first step of TIBET's bias discovery pipeline, which utilizes GPT 3.5 to come up with bias dimensions to be tested by the system. We observed that TIBET suggested generic dimensions, such as \emph{``age group''}, \emph{``occupation''}, or \emph{``appearance''} for S\textsubscript{gender}. However, it missed many of the experts' more specific observations, such as \emph{``happy''} or \emph{``bright colors''}. This might be because TIBET creates bias dimensions prior to the image generation step, just from the textual prompt.   
A full list of the automatically suggested bias dimensions can be found in Appendix~\ref{app:automatic}. We marked the test concepts that are covered by one of the bias axes from TIBET with a * in Table~\ref{tab:discoveredBiases}. Overall, for S\textsubscript{gender} and S\textsubscript{loc}, only 50\% of the test concepts found by the experts are also covered by TIBET.}

\section{Discussion and \rev{Conclusions}}
\label{sec:discussion}

Our expert study highlights that \name~successfully supports users discovering visual biases in T2I systems. \rev{Users reported SD3's bias towards \emph{``woman''} for the test concepts \emph{``beautiful''}, \emph{``dark skin''}, and \emph{``bright colors''} -- findings not previously documented in literature or via automated methods. We now revisit our design challenges} for further discussion and suggestions for future work. 

\rev{\emph{Flexibility through prompting tree:}} Our pilot study and the prompting trees constructed by the experts \rev{showed} that the prompting tree supports flexible expression of test concepts. However, users also faced difficulties correctly expressing their observed and suspected biases. In particular, defining and using appropriate relation types was a frequently observed difficulty. Here, we could illustrate that both CLIP and SD3 are, to a certain extent, resilient to grammar errors, and imperfect textual inputs may still lead to usable results (see Appendix~\ref{app:clip-robustness}). In the future, text parsing from the prompting tree could be enhanced by a language model to transform the parsed texts so that the respective model can interpret them more reliably. \rev{Such reformulations to less ambiguous or more well-understood concepts could potentially help reduce the presence of ``false positives'' where the user's intent of a test concept is not correctly interpreted by CLIP.}
\rev{In addition, as shown in Section~\ref{sec:validation}, some generic bias dimensions could easily be detected automatically. A future iteration of our workflow could, therefore, take a mixed-initiative approach where an automated system proposes and tests a large number of bias dimensions while the human user gets to test for more nuanced and specific bias dimensions.}

\rev{\emph{Instant zero-shot bias feedback:}} We showed that a bias measure based on image-text similarity in a multimodal embedding space can be computed within about a second. With server-side calculation and caching, zero-shot bias probing did not lead to any disruptive system lag.
In our case study, users explicitly appreciated the \emph{``instant feedback''} and confirmed that the quick bias estimations made the discovery process very engaging.

\rev{\emph{Improving zero-shot trustworthiness:}} Interactive bias exploration may introduce bias itself, for instance through a \emph{sample selection bias} by the image generator or a \emph{confirmation bias} by the user. In addition, the multimodal embedding space used for automatic zero-shot bias probing may be inherently biased. Our initial zero-shot bias probing performance experiments did not reveal any concerning results, but we occasionally observed that text input was not interpreted by CLIP as intended (see, for instance, Figure~\ref{fig:stripPlot} bottom) and CLIP indicating bias for test concepts that were in fact not depicted in the anchor concept images (e.g., \emph{``doctor''} and \emph{``nurse''}, as investigated by E2). 
\rev{To address transparency and ambiguity of vision-language models in the future, the data view could additionally show saliency maps~\cite{giulivi_concept_2024} of images on demand, thereby illustrating the image features that constitute the similarity to a selected test concept. If the user observes spurious correlations, they could then decide to exclude unrepresentative images from the zero-shot bias probing.}

\rev{\emph{Trustworthiness through confirmatory visualization:}} 
In the expert case study, we observed that all users consulted our provided visualizations for more in-depth confirmatory analysis. Similarly important are inverse bias queries to confirm the observed bias with a clear focus on the T2I model. For example, in the case of test concepts \emph{``doctor''} and \emph{``nurse''}, an a-posteriori inverse bias query clearly confirmed the existence of a gender bias in these two occupations for SD3. Unfortunately, inverse bias queries are slow due to the necessary image generation step. In the future, progressive image generation for faster approximate feedback could be a promising solution. \rev{This would allow for the image data to be delivered and processed in chunks, followed by updating the bias visualizations via progressive visual analytics strategies such as extension~\cite{fekete_progressive_2024, ulmer_survey_2024}.}

\rev{\emph{Scalability:} A challenge not explicitly addressed in our reference implementation is scalability with respect to the number of sample images and anchor concepts. The \name~workflow and zero-shot bias probing conceptually allow for an extension to an arbitrary number of anchor concepts and sample images. However, the visualizations will have to be adjusted: Both the strip plot and inverse bias scatterplot are trivially scalable. Here, to reduce clutter, a kernel density plot could be used instead of showing the images directly. In the prompting tree an option would be to employ pair-wise color codings and only show the similarity values between the top two anchor concepts per node. For the inverse bias scatterplot, a radar chart or a parallel coordinates representation with an axis per anchor concept would be suitable candidates.}

\rev{\emph{Generalizability:}} In this work, we employed \name~for SD3 and a subset of the FairFace image collection. In the future, we hope to see \name~in experiments comparing T2I models trained on different data sets, as well as original and de-biased models.

\rev{In conclusion, with \name, we demonstrated how users can flexibly explore visual bias in T2I models at a rapid pace while maintaining trust through confirmatory visualizations. We showed that, with our proposed workflow, users quickly can come up with new bias dimension candidates. We see the greatest potential for future improvements in mixed-initiative bias queries, explainable zero-shot bias probes, and faster inverse bias queries. }
\rev{Beyond exploration and confirmation of visual bias, \name~could ultimately be used for finding bias mitigation strategies by using zero-shot bias probing to create efficient prompt injections.}

\rev{\section*{Acknowledgments}
We thank Dominik Wolf for his contributions to the software framework and the participants of the studies. This research was funded in whole or in part by the Austrian Science Fund (FWF) \href{https://doi.org/10.55776/P36453}{10.55776/P36453} and by the Austrian Research Promotion Agency (FFG), project no. 898085 and FO999904624.}

\bibliographystyle{eg-alpha-doi} 
\bibliography{references-static}  

\newpage
\appendix

\section{Prompt Sensitivity}
\label{app:clip-robustness}

\begin{figure}[H]
    \centering
    \begin{subfigure}[]{\linewidth}
        \centering
        \includegraphics[width=\textwidth]{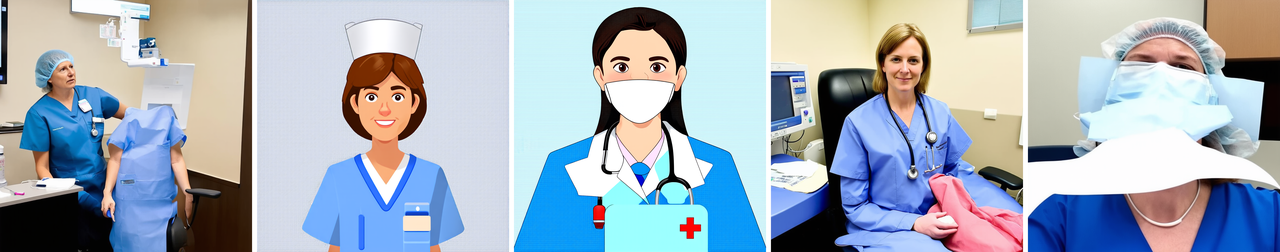}
        \caption{Prompt: \emph{``picture that shows a person that shows a nurse''}.}
        \label{fig:SD3-nurse-wrong}
    \end{subfigure}
    \hfill
    \begin{subfigure}[]{\linewidth}
        \centering
        \includegraphics[width=\textwidth]{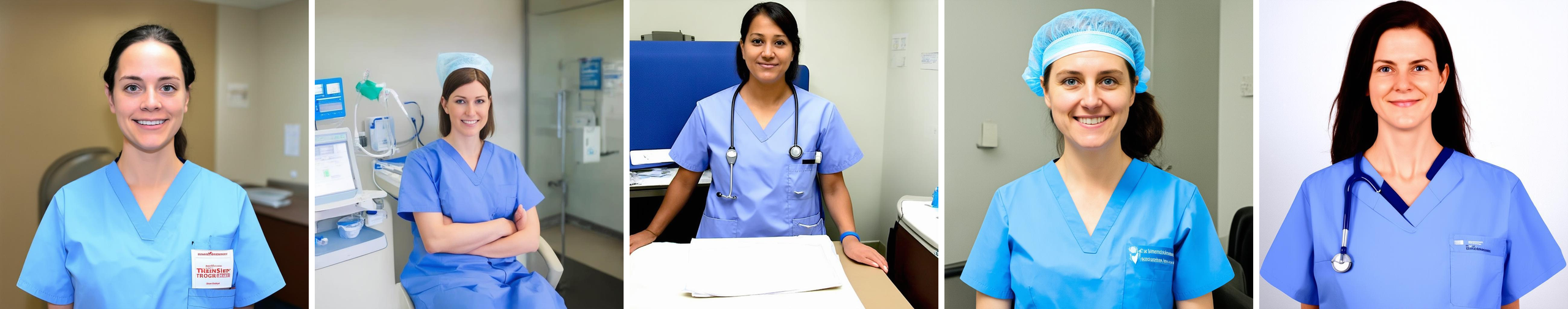}
        \caption{Prompt: \emph{``picture that shows a nurse''}.}
        \label{fig:SD3-nurse-correct}
    \end{subfigure}
    \caption{Images generated with SD3 comparing two prompt formulations for \emph{``nurse''}.}
    \label{fig:SD3-nurse}
\end{figure}

\begin{figure}[H]
    \centering
    \begin{subfigure}[]{\linewidth}
        \centering
        \includegraphics[width=\textwidth]{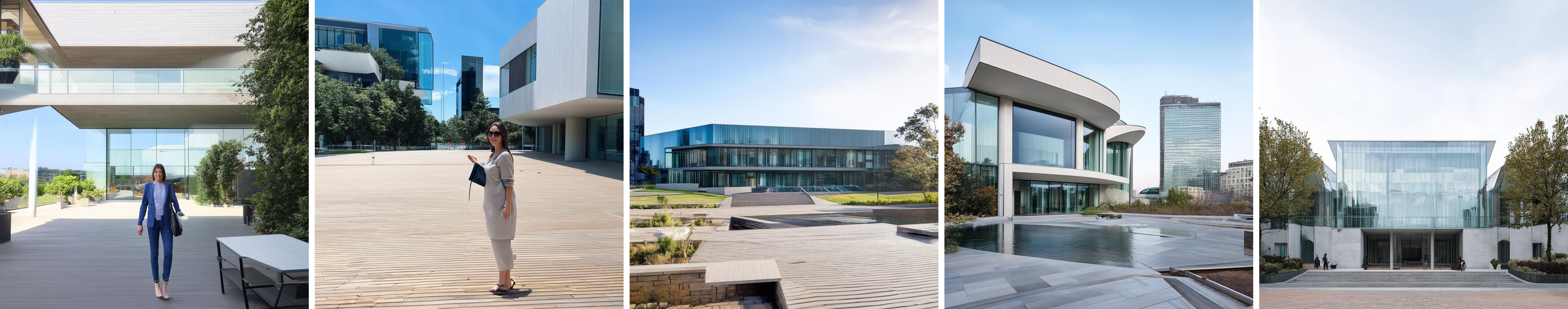}
        \caption{Prompt: \emph{``picture with modern architecture focus content''}.}
        \label{fig:SD3-architecture-wrong}
    \end{subfigure}
    \hfill
    \begin{subfigure}[]{\linewidth}
        \centering
        \includegraphics[width=\textwidth]{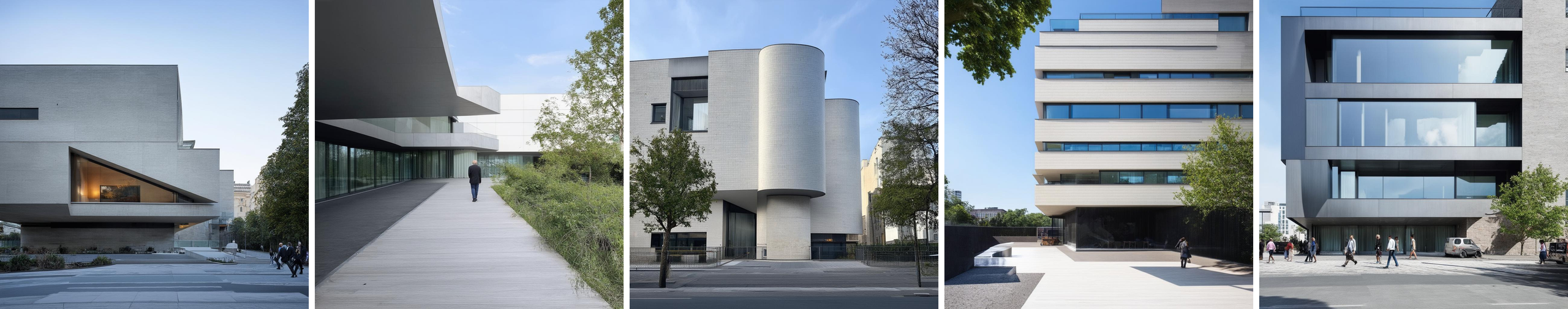}
        \caption{Prompt: \emph{``picture that shows modern architecture''}.}
        \label{fig:SD3-architecture-correct}
    \end{subfigure}
    \caption{Images generated with SD3 comparing two prompt formulations for \emph{``modern architecture''}.}
    \label{fig:SD3-architecture}
\end{figure}

\begin{figure}[H]
    \centering
    \begin{subfigure}[]{\linewidth}
        \centering
        \includegraphics[width=\textwidth]{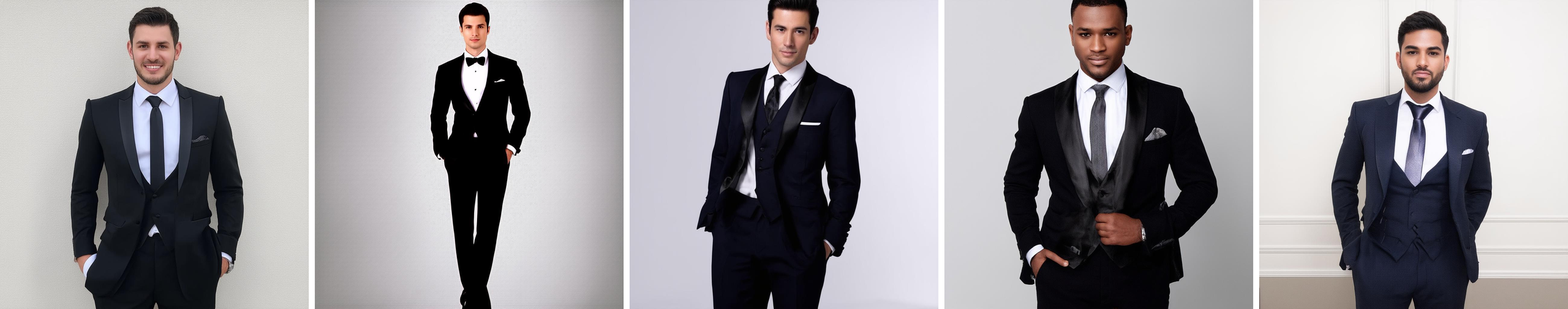}
        \caption{Prompt: \emph{``picture of a formal wear person''}.}
        \label{fig:SD3-formal-wrong}
    \end{subfigure}
    \hfill
    \begin{subfigure}[]{\linewidth}
        \centering
        \includegraphics[width=\textwidth]{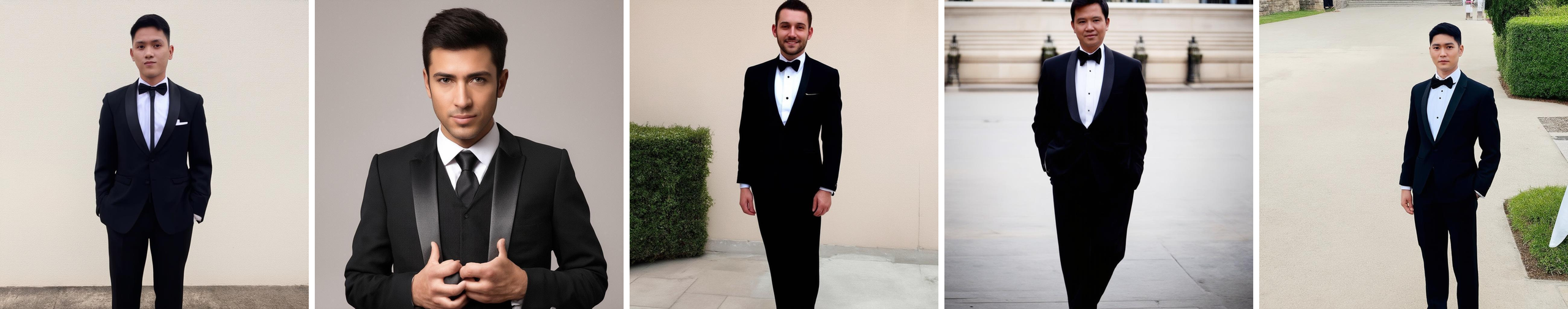}
        \caption{Prompt: \emph{``picture that shows a person with formal wear''}.}
        \label{fig:SD3-formal-correct}
    \end{subfigure}
    \caption{Images generated with SD3 comparing two prompt formulations for \emph{``formal wear''}.}
    \label{fig:SD3-formal}
\end{figure}

Our evaluation showed that the prompting tree may lead to malformed prompts due to inaccurate relations. The literature shows that T2I models emphasize \emph{keywords} (i.e., nodes in the prompting tree) over \emph{connecting words} (edges)~\cite{liu_design_2022}. We performed tests prompting SD3 with malformed prompts from both the prompting tree study and the expert case study. Results of these tests indicate that inaccurate relations do lead to the same image content but may decrease image quality (Figure~\ref{fig:SD3-nurse} - \ref{fig:SD3-formal}).

\begin{figure}[H]
    \centering
    \includegraphics[width=\linewidth]{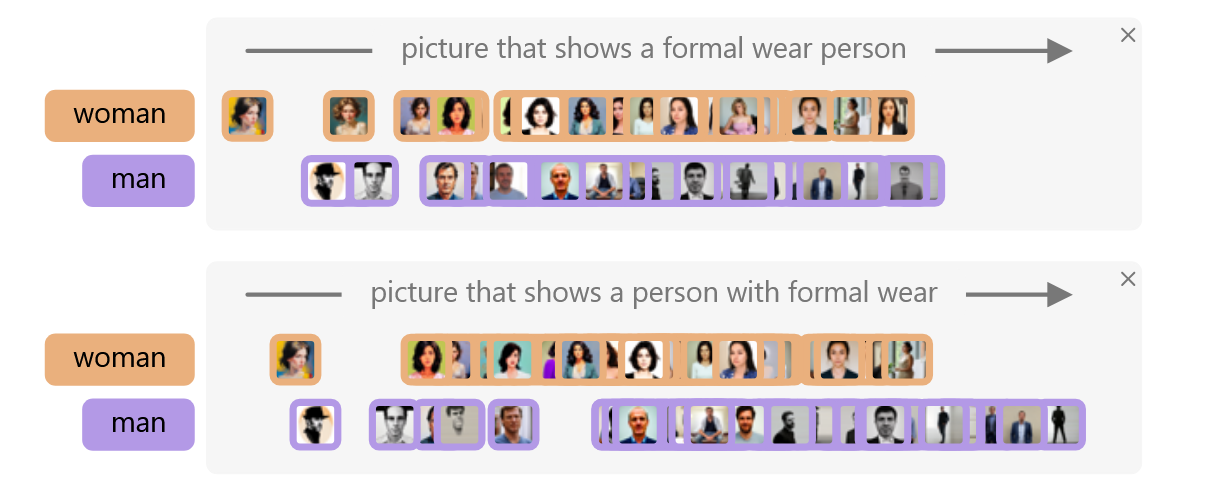}
    \caption{Distribution of similarities for \emph{``formal wear person''} vs. \emph{``person with formal wear''} in S\textsubscript{gender}.}
    \label{fig:clip-robustness-formal}
\end{figure}

\begin{figure}[H]
    \centering
    \includegraphics[width=\linewidth]{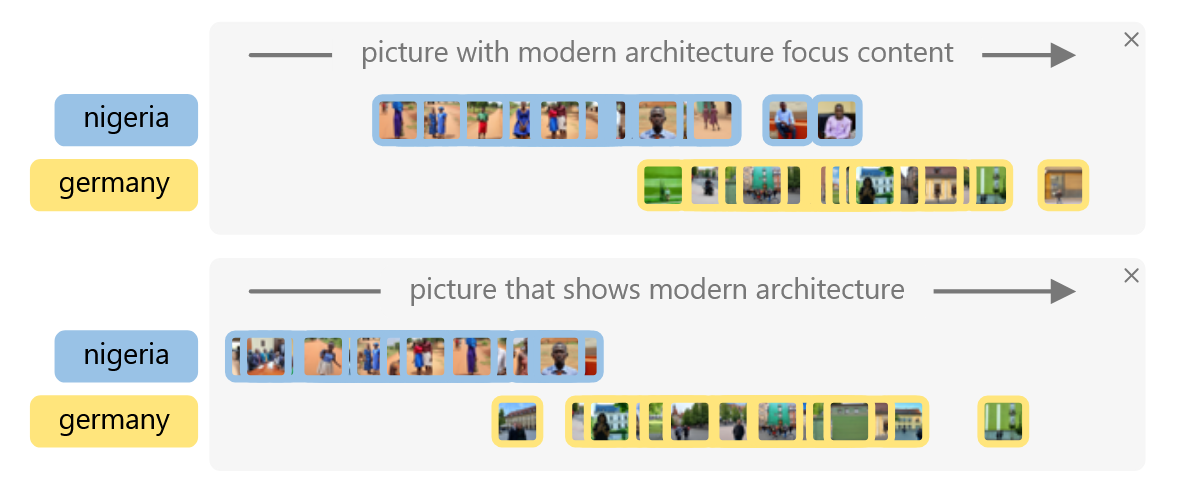}
    \caption{Distribution of similarities for \emph{``modern architecture focus content''} vs. \emph{``modern architecture''} in S\textsubscript{loc}.}
    \label{fig:clip-robustness-architecture}
\end{figure}

Furthermore, we also computed CLIP similarity scores with malformed prompts to see how bias queries are affected (Figure~\ref{fig:clip-robustness-formal} and \ref{fig:clip-robustness-architecture}). Here we observe similar behavior as with SD3, where the general distributions are comparable, but separations are less pronounced. As these tests are only of an exploratory nature, a more thorough evaluation remains to be conducted in future work. 

\begin{figure*}[ht]
    \centering
    \includegraphics[width=\linewidth]{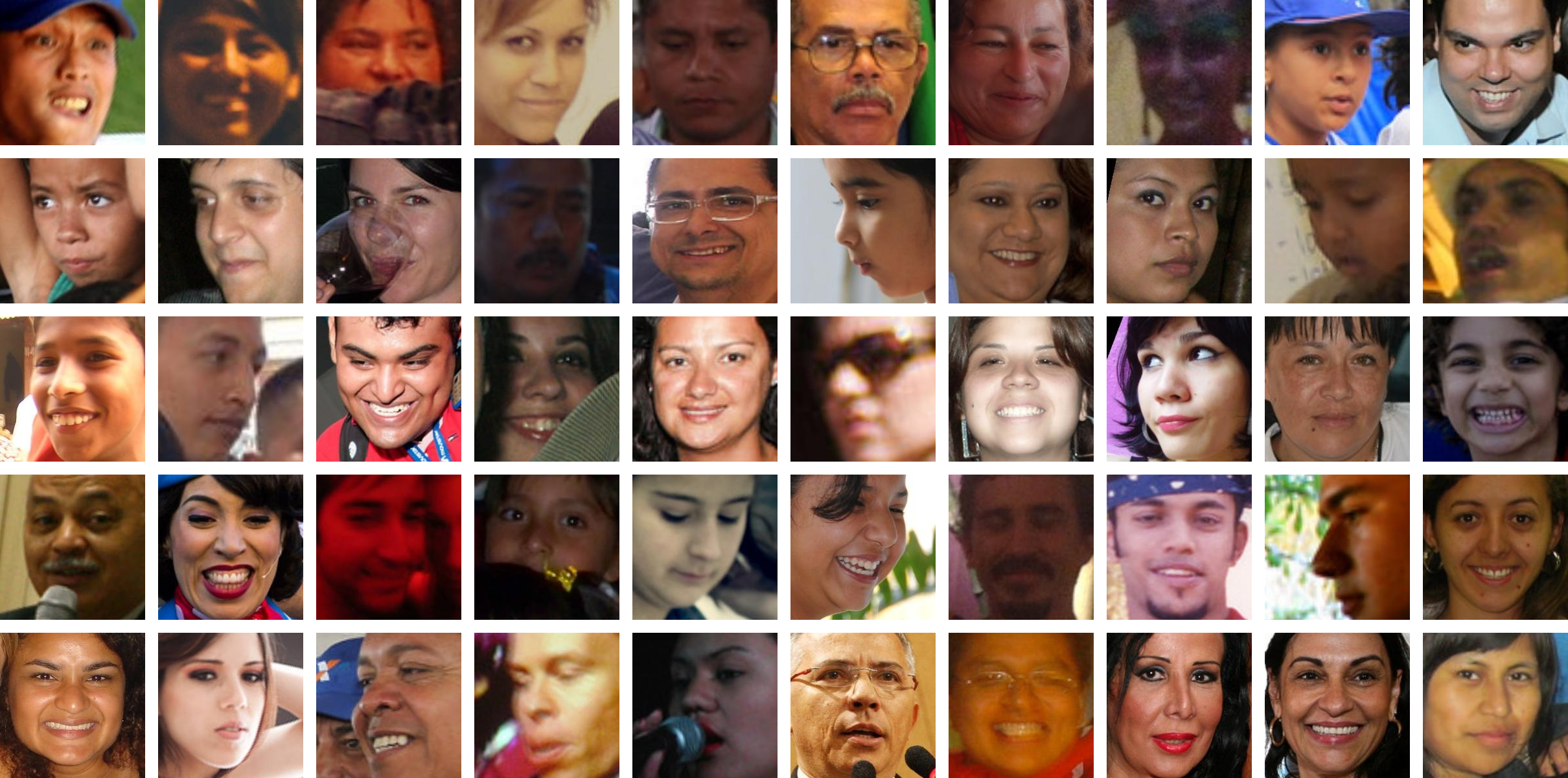}
    \caption{Gender- and age-balanced sample from FairFace for race \emph{``latino''}.}
    \label{fig:ff-latino}
\end{figure*}

\begin{figure*}[ht]
    \centering
    \includegraphics[width=\linewidth]{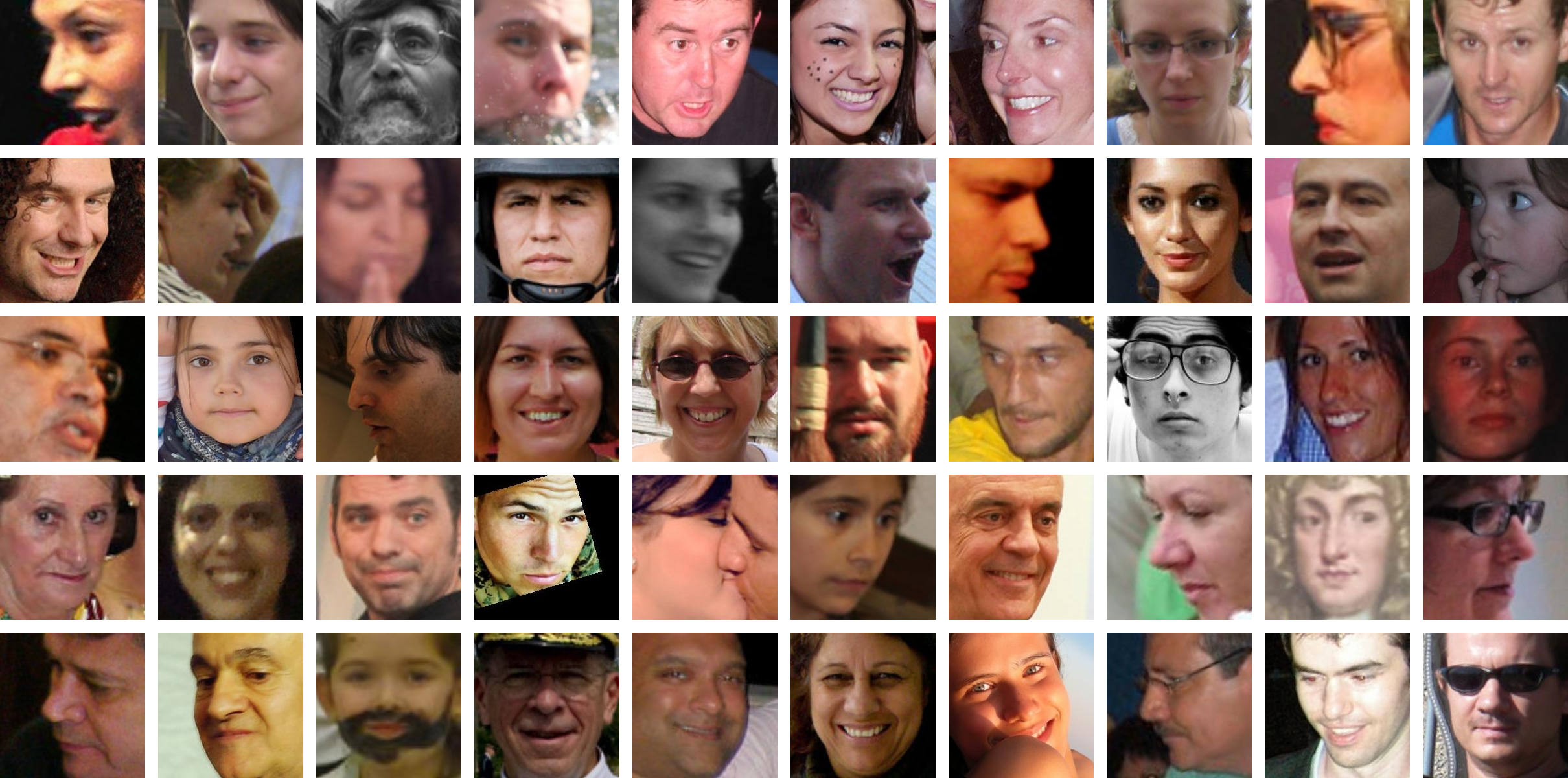}
    \caption{Gender- and age-balanced sample from FairFace for race \emph{``white''}.}
    \label{fig:ff-white}
\end{figure*}

\rev{
\section{CLIP Experiments}
\label{app:clip-experiments}
We systematically compared the similarity measures produced by CLIP (see Equation~\ref{eq:similarity}) for test concepts with known ground truth to validate the reliability of CLIP for zero-shot bias probing (\textbf{\rev{C3}}).
We conducted forward bias queries for two scenarios: S\textsubscript{gender} against the test concept \emph{``beard''} and S\textsubscript{race} against the test concept \emph{``person''}. The first scenario thereby represents an expected entanglement, where we expect to see a tendency towards \emph{``man''}. In the second scenario, \emph{``person''} should be balanced between \emph{``Latino''} and \emph{``Caucasian''}. 50 images per anchor concept were manually selected from FairFace~\cite{karkkainen_fairface_2021} (see Figures~\ref{fig:ff-latino} and~\ref{fig:ff-white}). For the second experiment, we measured the response times for similarity computations of ten different test concepts taken from the prompting tree pilot study and 100 anchor images from S\textsubscript{loc}. 

\paragraph*{Results:} For the scenario S\textsubscript{gender}, a Kolmogorov-Smirnov test showed a significant difference between the similarity distributions ($D(50)=.653, p<.001$). For the scenario S\textsubscript{race}, we do not see a significant difference ($D(50)=.16, p=.548$). This confirms that CLIP can correctly detect an expected and obvious entanglement (i.e., men are more associated with beards than women), while it shows no significant difference for a balanced image distribution (i.e., an equal number of Latino and Caucasian persons). 
}

\rev{\section{Discovered Biases}
\label{app:discovered-biases}

Figure~\ref{fig:FFEval} shows the results of the bias validation via FairFace~\cite{karkkainen_fairface_2021}. Here, we plot the absolute frequency of the two anchor concepts for each test concept.

\begin{figure*}[htbp!]
    \centering
    \includegraphics[width=\linewidth]{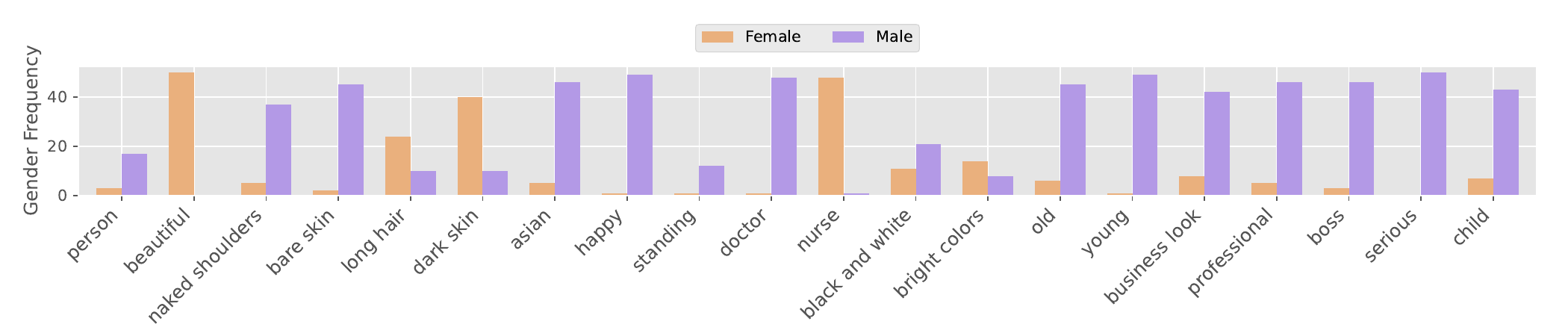}
    \caption{Results of the bias validation. For each bias from S\textsubscript{gender} listed in Table~\ref{tab:discoveredBiases}, we generated 50 new images and plot the frequency of \emph{``male''} and \emph{``female''} as classified by FairFace. Note that FairFace is only able to classify those images with visible faces.}
    \label{fig:FFEval}
\end{figure*}
}

\rev{\section{Automated Bias Evaluation Experiment}
\label{app:automatic}

Tables~\ref{tab:tibet-gender} and~\ref{tab:tibet-loc} show the bias dimensions (formulated as questions) for scenarios S\textsubscript{gender} and S\textsubscript{loc} obtained from running the bias axis generation step of TIBET~\cite{chinchure_tibet_2024}.

\begin{table*}[htbp!]
\centering
\begin{tabular}{ll}
\hline
\multicolumn{2}{c}{\textbf{S\textsubscript{gender}}}                        \\ \hline
picture that shows a woman                                              & picture that shows a man                        \\ \hline
What is the gender (male, female, other) of the person in the image?    & What is the gender (male, female, other) of the person in the image? \\
What is cultural in the image?                                          & Describe the appearance in the image.                           \\
What age group (young, middle, old) is the person in the image?         & Describe the background of the image.                           \\
Describe the appearance in the image.                                   & What is the occupation of the person in the image?              \\
What is societal in the image?                                          & What age group (young, middle, old) is the person in the image? \\
                                                                        & What is body type in the image?                                 \\ \hline
\end{tabular}

\caption{Automatically generated bias dimensions by TIBET~\cite{chinchure_tibet_2024} for scenario S\textsubscript{gender}.}
\label{tab:tibet-gender}
\end{table*}

\begin{table*}[htbp!]
\centering
\begin{tabular}{ll}
\hline
\multicolumn{2}{c}{\textbf{S\textsubscript{loc}}}                        \\ \hline
picture taken in germany         & picture taken in nigeria              \\ \hline
What is geographic in the image? & What is geographic in the image?      \\
What is cultural in the image?   & What is cultural in the image?        \\
What is historical in the image? & What is economic in the image?        \\
What is touristic in the image?  & What is urban vs. rural in the image? \\
                                 & What is political in the image?       \\
                                 & What is historical in the image?      \\ \hline
\end{tabular}

\caption{Automatically generated bias dimensions by TIBET~\cite{chinchure_tibet_2024} for scenario S\textsubscript{loc}.}
\label{tab:tibet-loc}
\end{table*}

}

\section{Bias Exploration Scenarios}
\label{app:biasScenarios}
\begin{figure*}[ht]
    \centering
    \includegraphics[width=\linewidth]{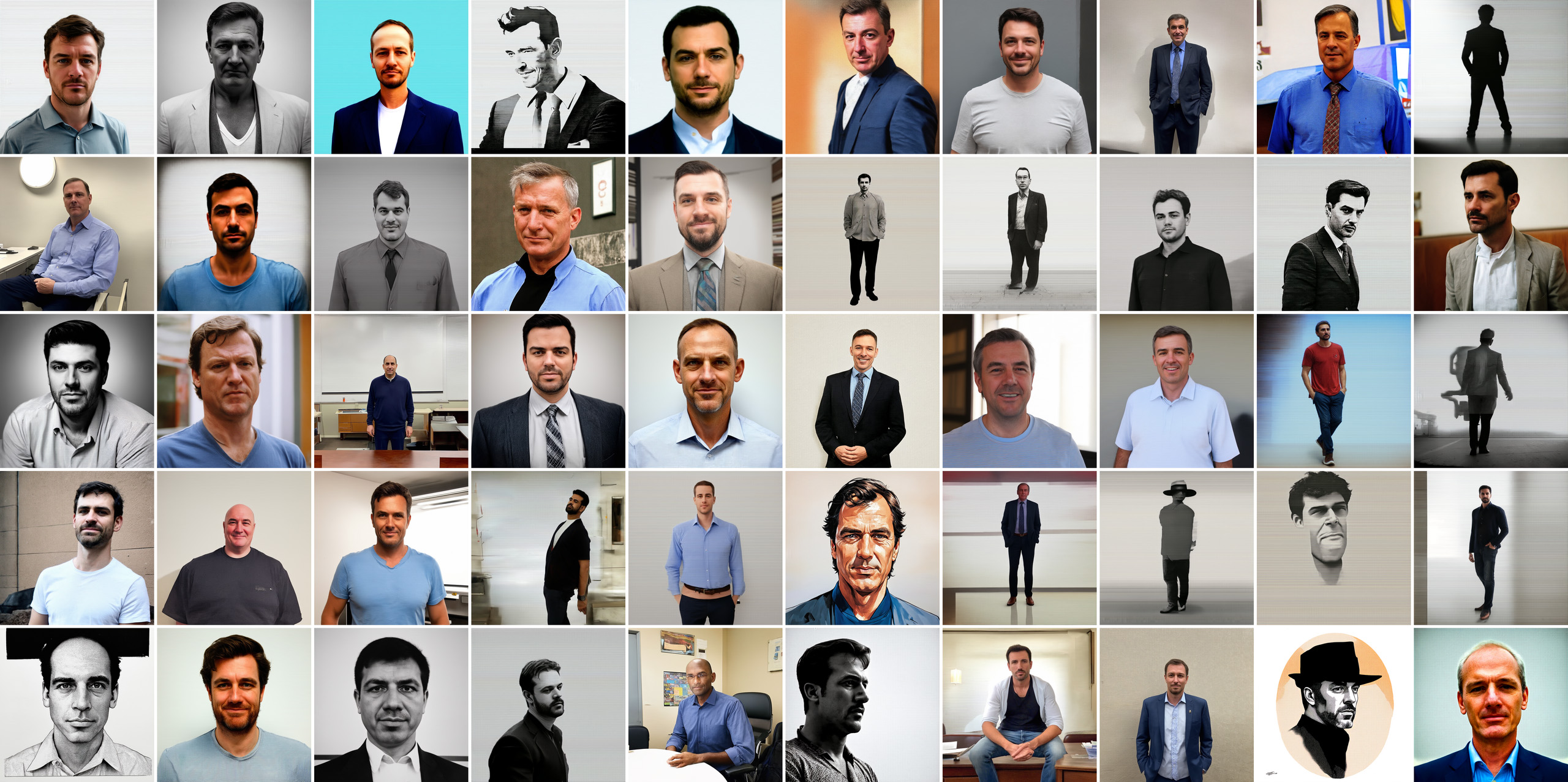}
    \caption{S\textsubscript{gender}. Prompt: \emph{``picture that shows a man''}.}
    \label{fig:sgender-man}
\end{figure*}

\begin{figure*}[ht]
    \centering
    \includegraphics[width=\linewidth]{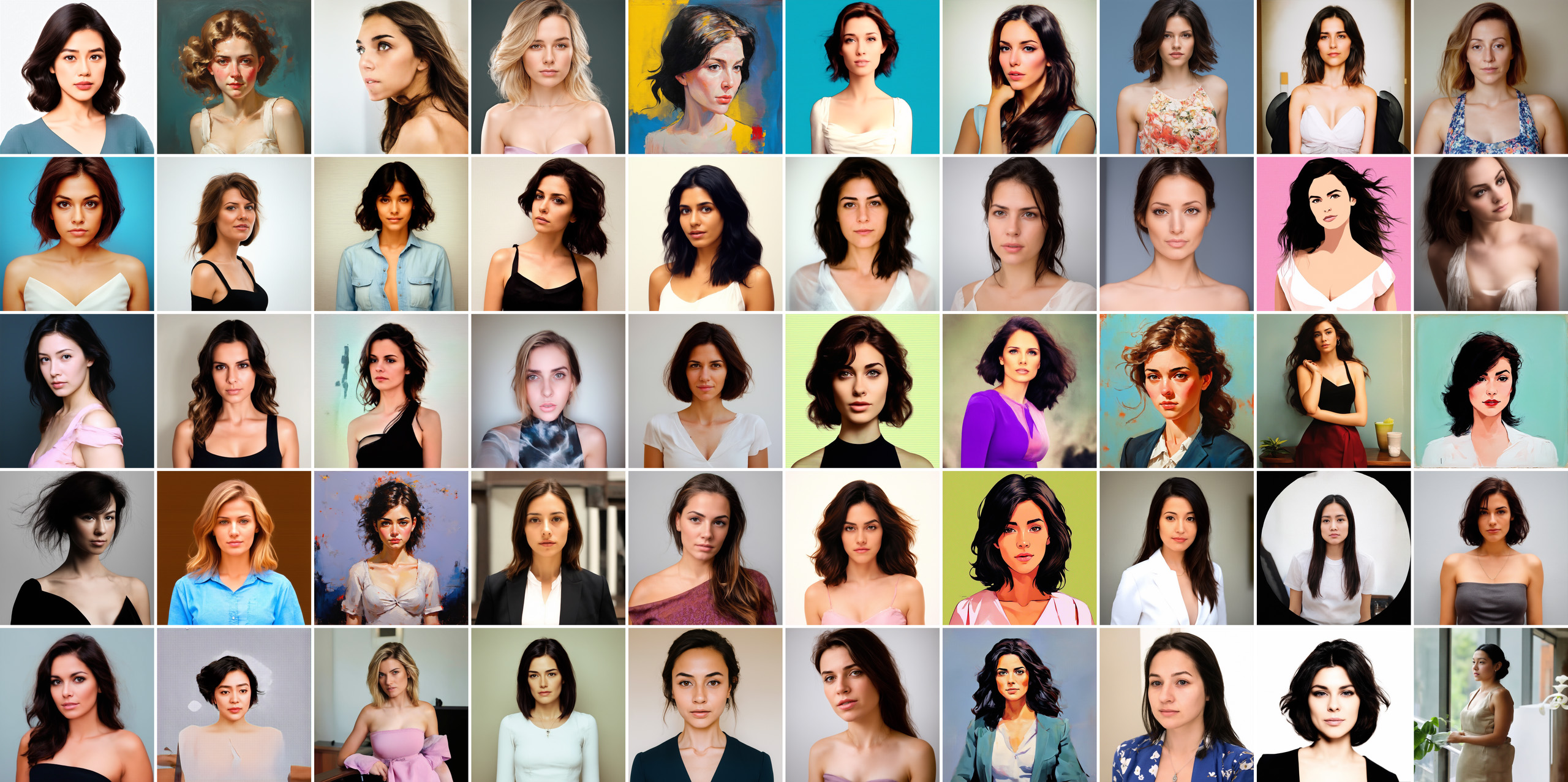}
    \caption{S\textsubscript{gender}. Prompt: \emph{``picture that shows a woman''}.}
    \label{fig:sgender-woman}
\end{figure*}

\begin{figure*}[ht]
    \centering
    \includegraphics[width=\linewidth]{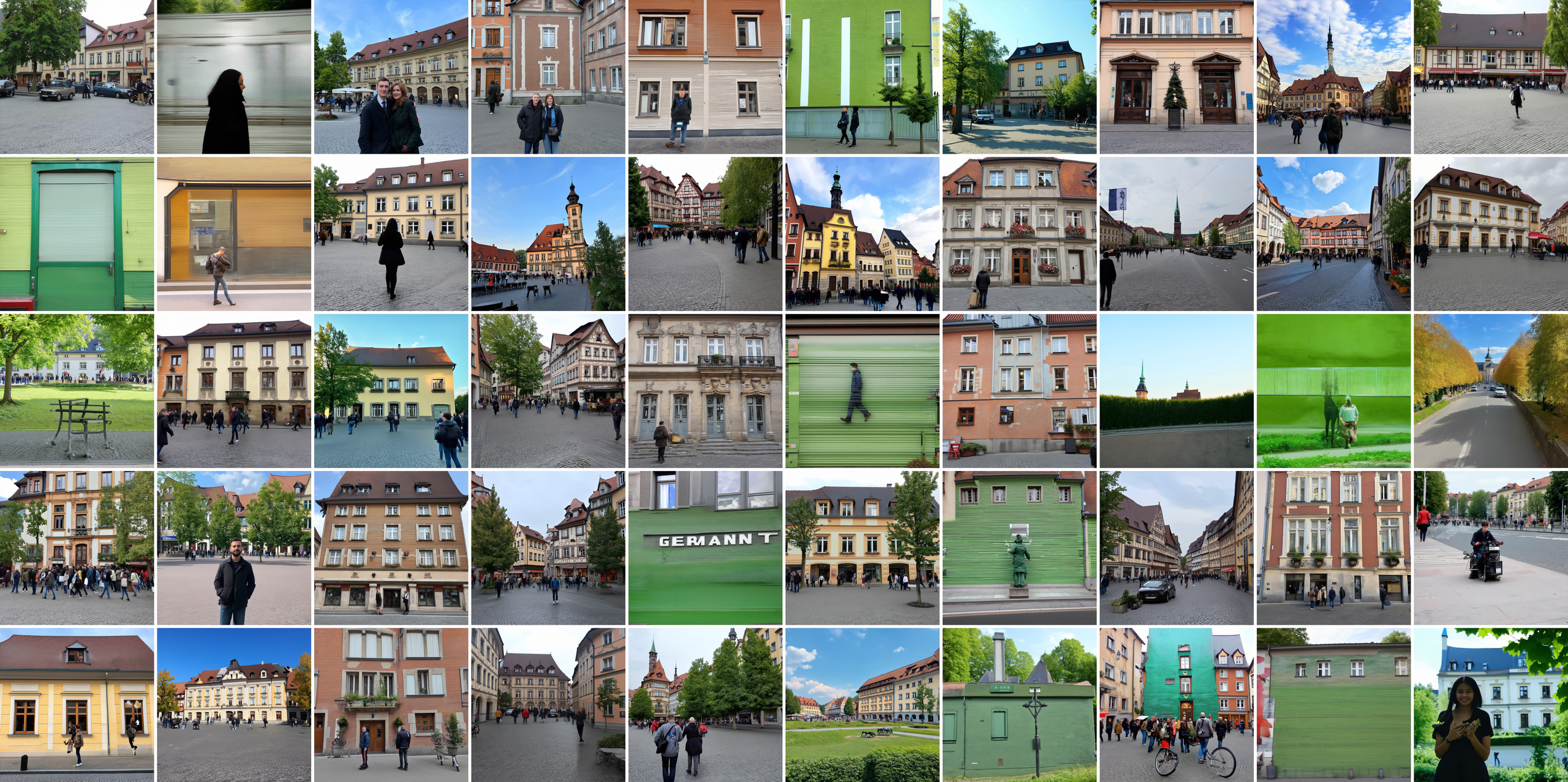}
    \caption{S\textsubscript{loc}. Prompt: \emph{``picture taken in Germany''}.}
    \label{fig:sloc-germany}
\end{figure*}

\begin{figure*}[ht]
    \centering
    \includegraphics[width=\linewidth]{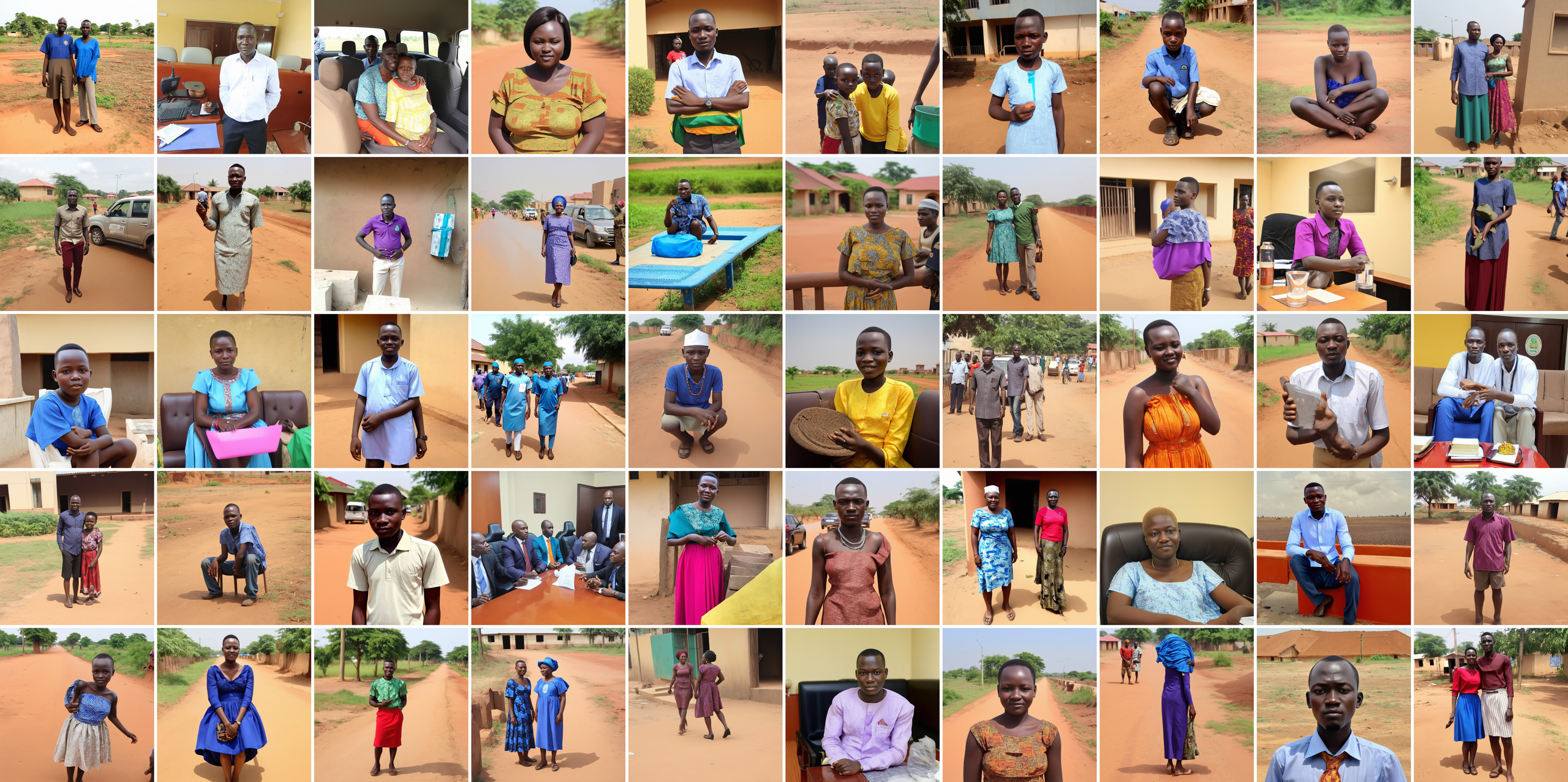}
    \caption{S\textsubscript{loc}. Prompt: \emph{``picture taken in Nigeria''}.}
    \label{fig:sloc-nigeria}
\end{figure*}

\begin{figure*}[ht]
    \centering
    \includegraphics[width=\linewidth]{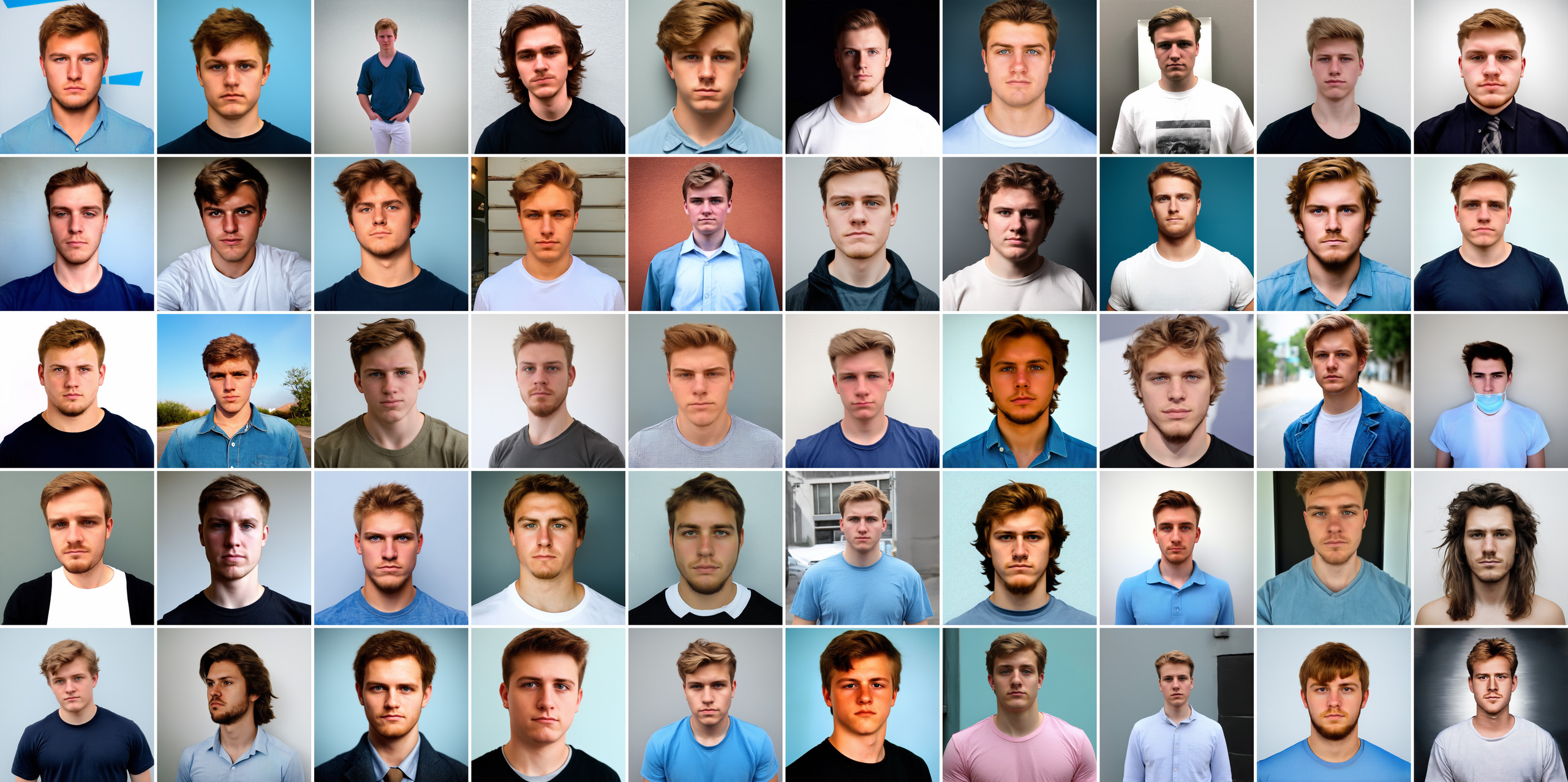}
    \caption{S\textsubscript{race}. Prompt: \emph{``picture of a caucasian person''}.}
    \label{fig:srace-caucasian}
\end{figure*}

\begin{figure*}[ht]
    \centering
    \includegraphics[width=\linewidth]{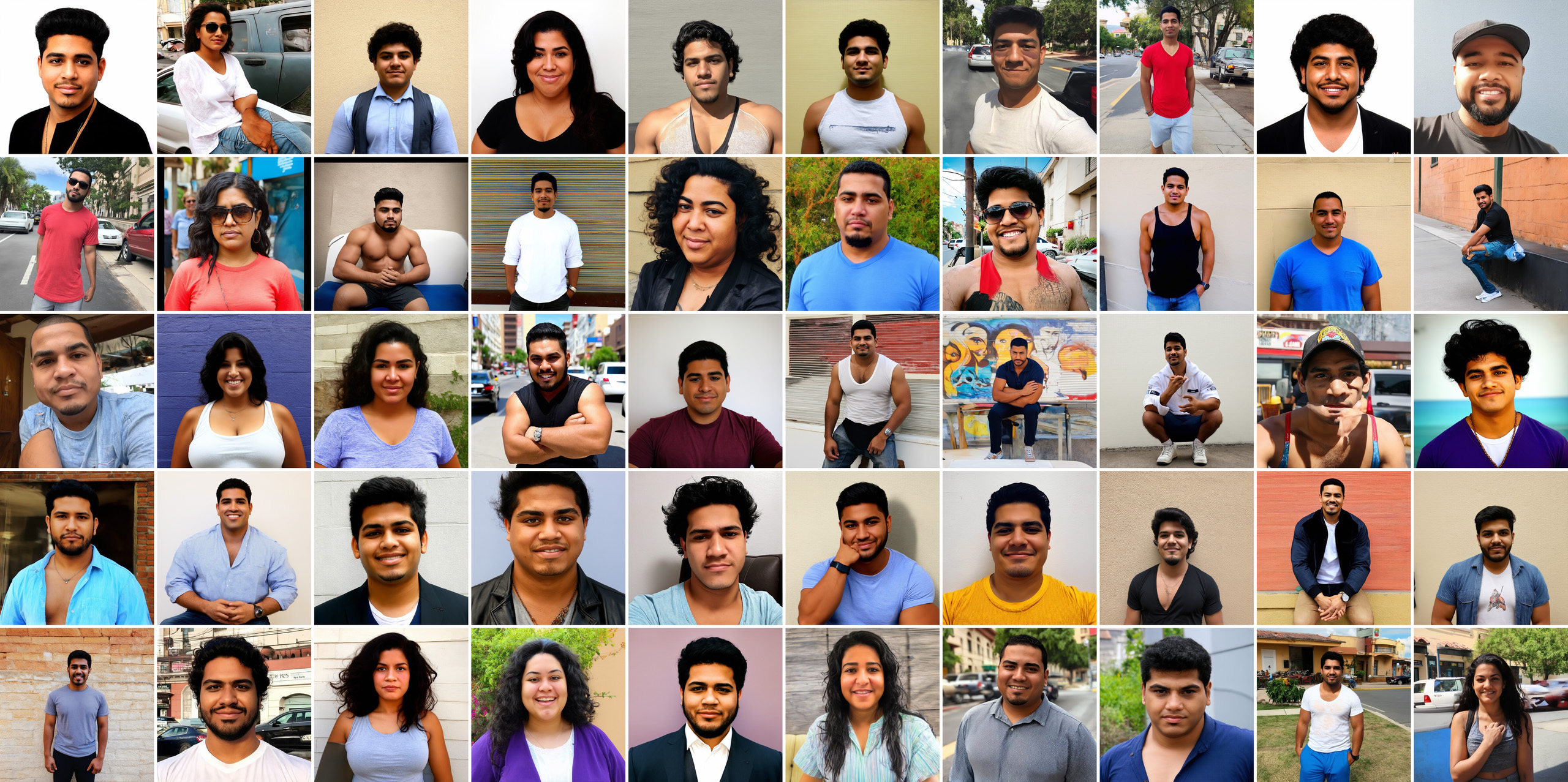}
    \caption{S\textsubscript{race}. Prompt: \emph{``picture of a latino person''}.}
    \label{fig:srace-latino}
\end{figure*}

Figures~\ref{fig:sgender-man} - \ref{fig:srace-latino} show all images used for our three bias exploration scenarios.

\section{Expert Case Study Prompting Trees}
\label{app:expertTrees}

\begin{figure*}[ht!]
    \centering
    \begin{subfigure}[]{0.5\linewidth}
        \centering
        \includegraphics[width=\textwidth]{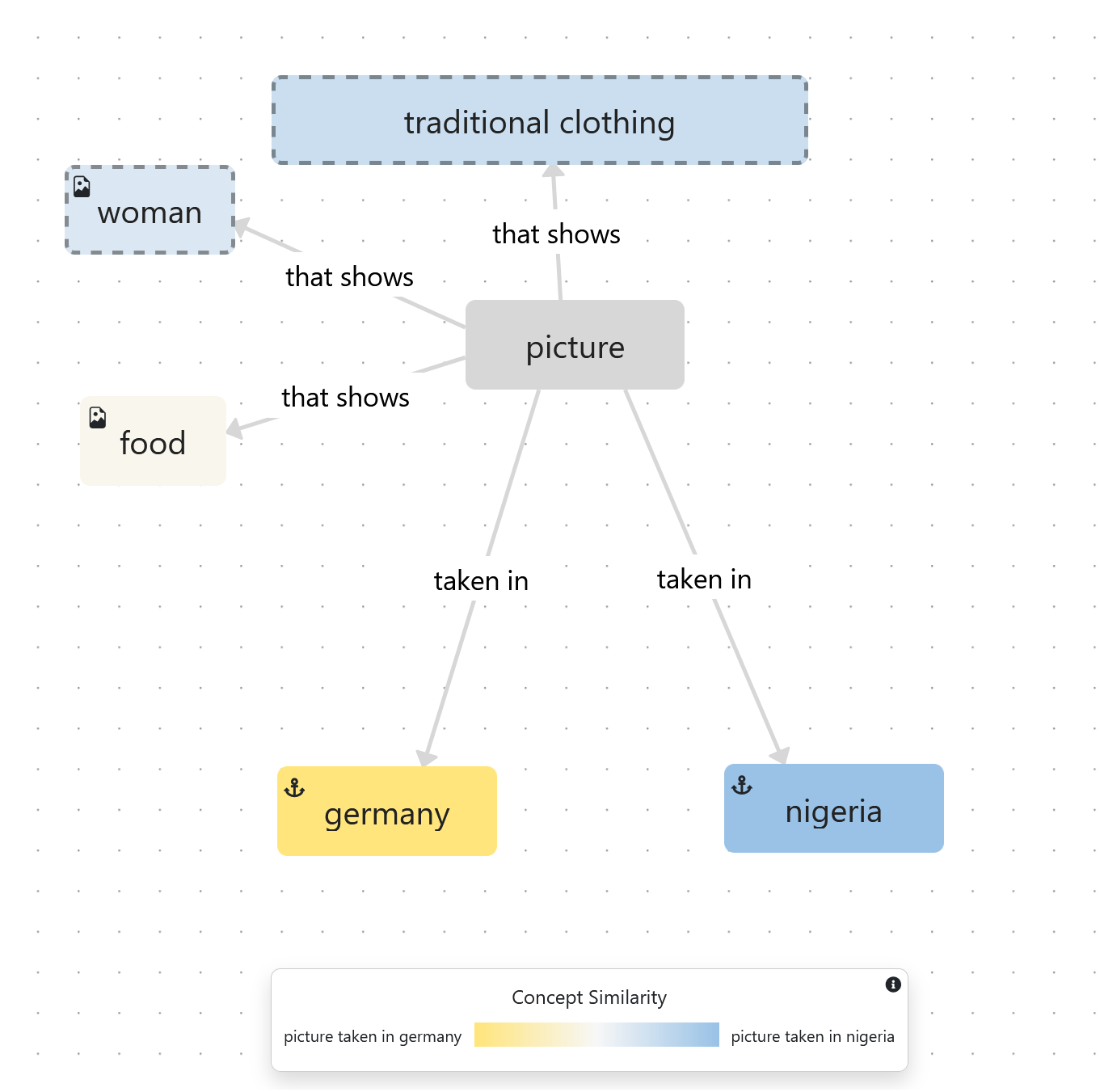}
        \caption{E3}
        \label{fig:s1e03}
    \end{subfigure}
    \hfill
    \begin{subfigure}[]{0.75\linewidth}
        \centering
        \includegraphics[width=\textwidth]{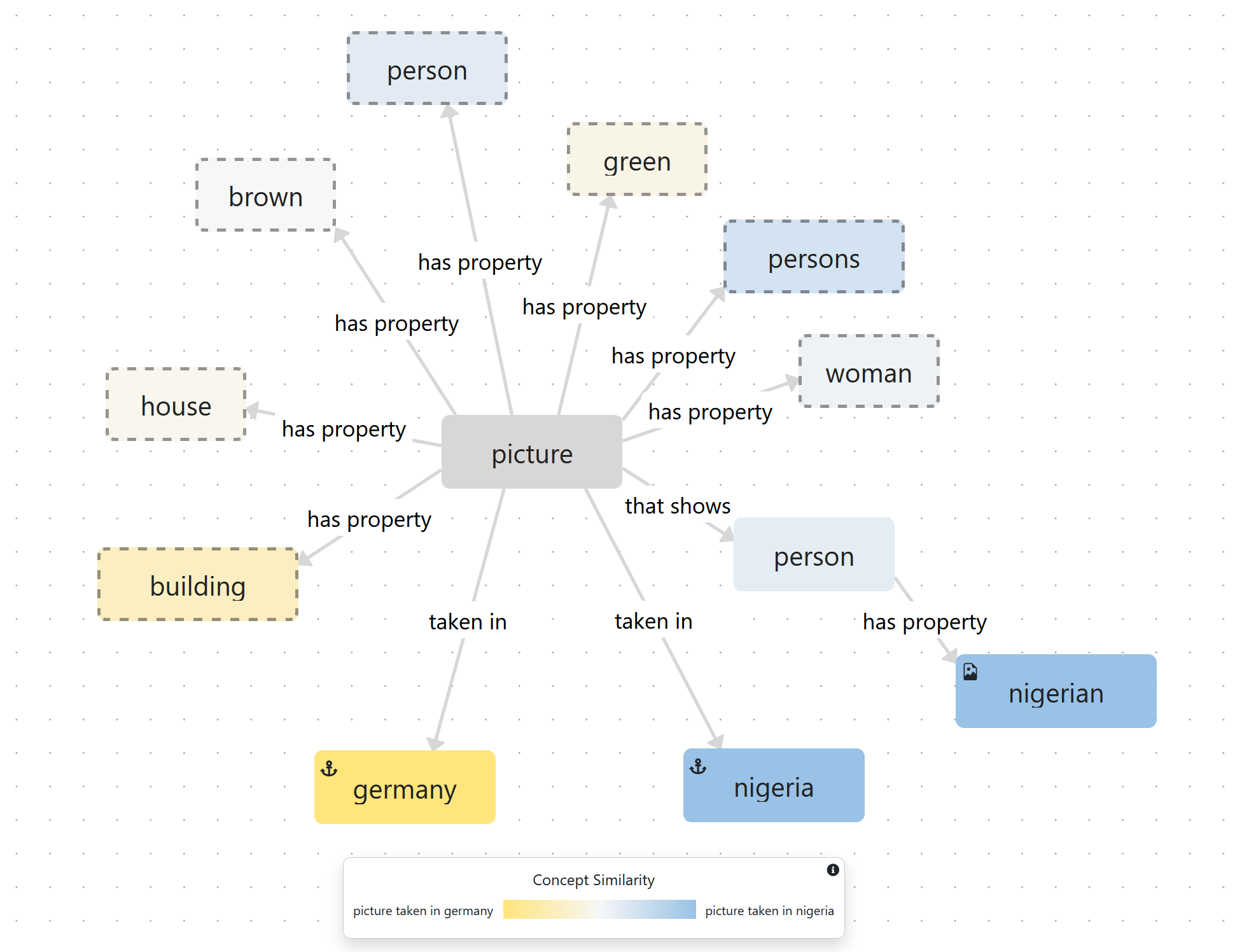}
        \caption{E4}
        \label{fig:s1s04}
    \end{subfigure}

    \caption{Prompting trees generated during the case studies for S\textsubscript{loc}.}
    \label{fig:s1-trees}
\end{figure*}

\begin{figure*}[ht!]
    \centering
    \begin{subfigure}[]{0.48\linewidth}
        \centering
        \includegraphics[width=\textwidth]{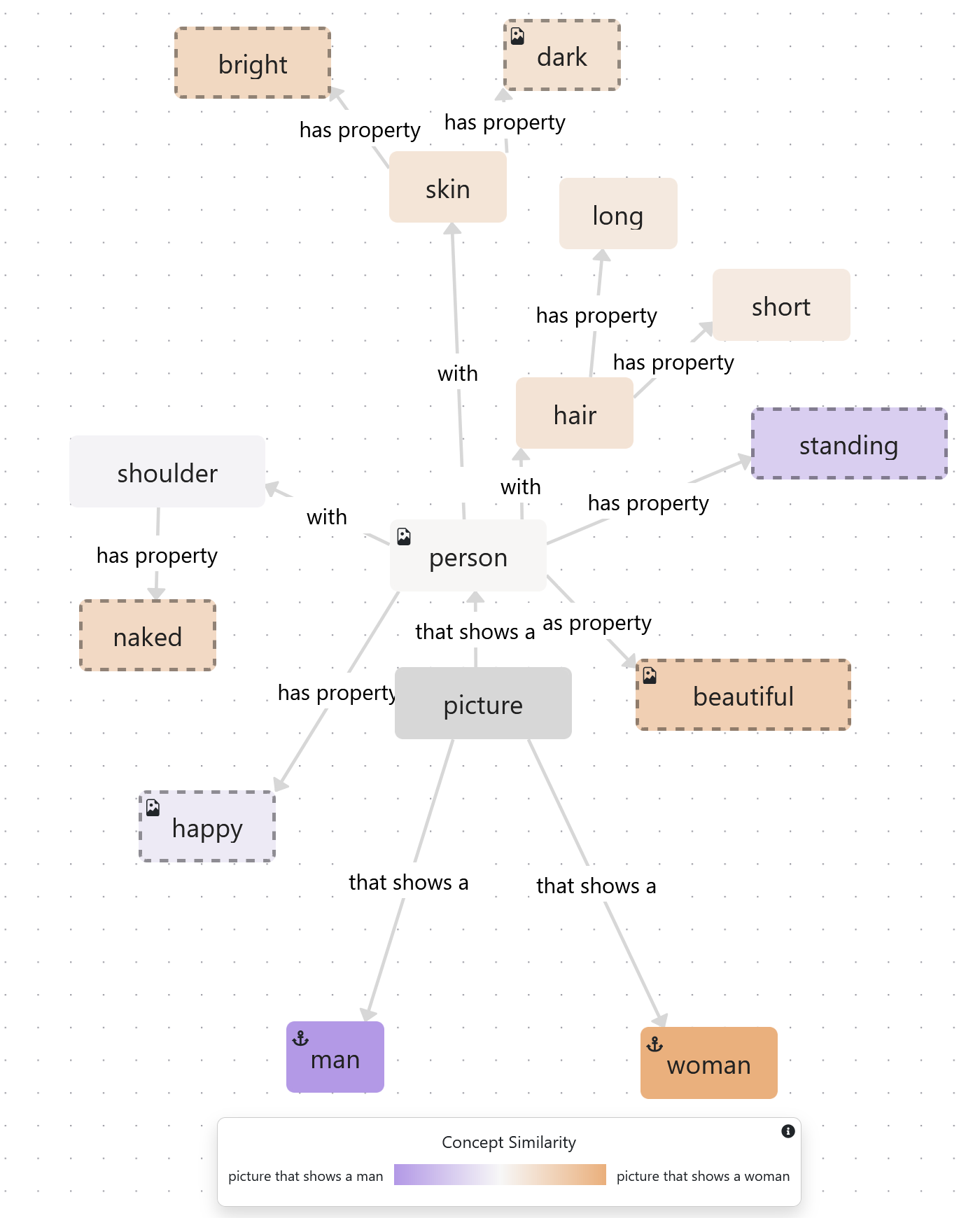}
        \caption{E1}
        \label{fig:s2e01}
    \end{subfigure}
    \hfill
    \begin{subfigure}[]{0.48\linewidth}
        \centering
        \includegraphics[width=\textwidth]{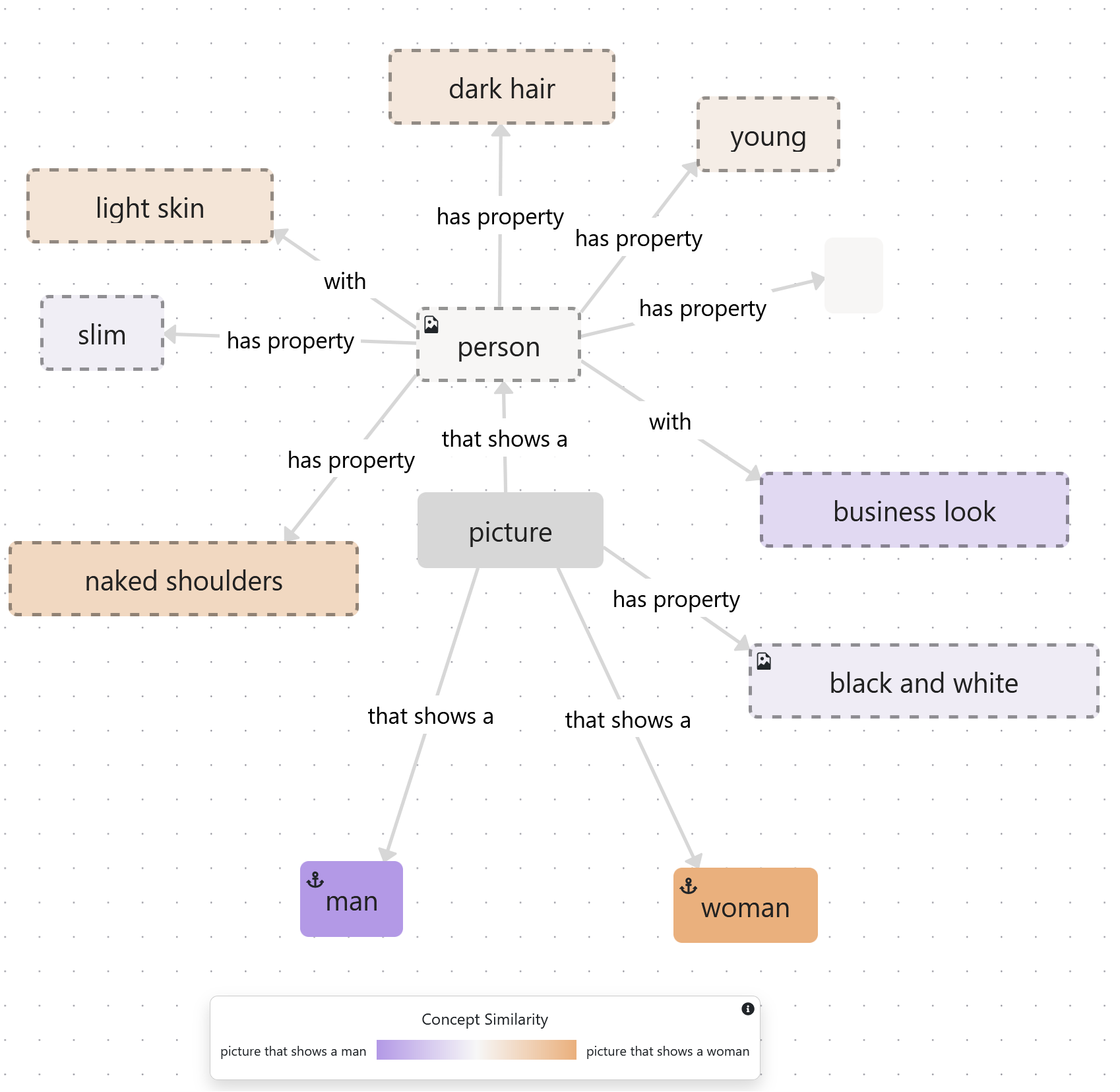}
        \caption{E4}
        \label{fig:s2e04}
    \end{subfigure}

    \vspace{1em}

     \begin{subfigure}[]{0.72\linewidth}
        \centering
        \includegraphics[width=\textwidth]{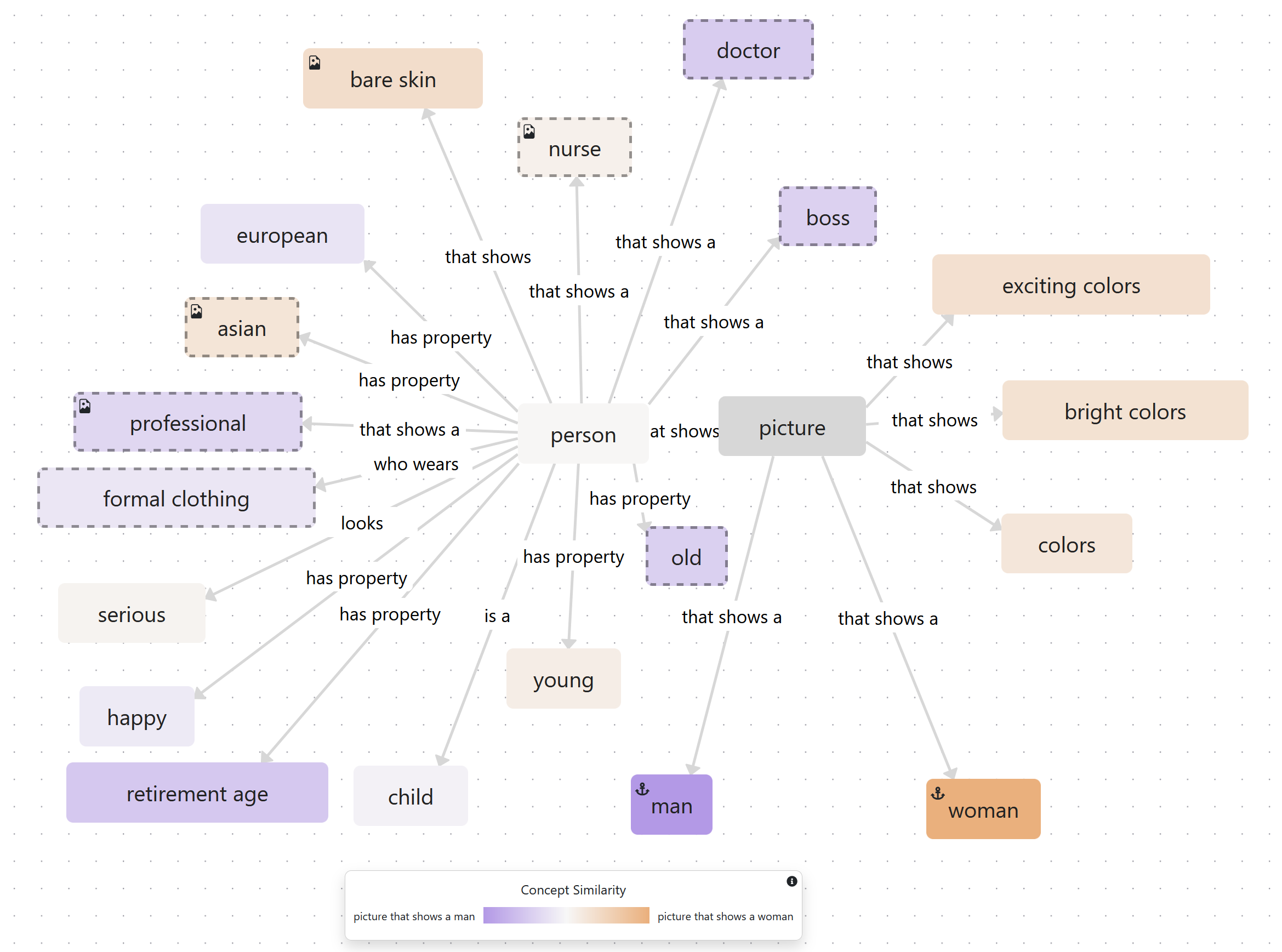}
        \caption{E2}
        \label{fig:s2e02}
    \end{subfigure}

    \caption{Prompting trees generated during the case studies for S\textsubscript{gender}.}
    \label{fig:s2-trees}
\end{figure*}

Figures~\ref{fig:s1-trees} and \ref{fig:s2-trees} show all prompting trees authored during the expert case study.

\end{document}